\documentclass[%
reprint,
superscriptaddress,
 amsmath,amssymb,
 aps,
 prd,
 numerical,
 longbibliography,
]{revtex4-2}

\usepackage{graphicx}
\usepackage{dcolumn}
\usepackage{bm}

\usepackage[varg]{txfonts}
\usepackage[T1]{fontenc}
\usepackage[utf8]{inputenc}   
\DeclareUnicodeCharacter{2243}{\ensuremath{\simeq}}
\usepackage[colorlinks=true, allcolors=blue]{hyperref}
\usepackage{xcolor}
\usepackage{float}
\usepackage{aas_macros_prd}
\usepackage[switch]{lineno}

\defcitealias{Lelli2017}{L17}
\defcitealias{Kelleher:2024oip}{KL24}

\begin{document}

\title{Revisiting the missing mass problem in MOND for nearby galaxy clusters}

\author{Dong Zhang}
\email{zhang\_dong\_96@163.com}
\affiliation{Helmholtz-Institut für Strahlen-und Kernphysik (HISKP), Universität Bonn, Nussallee 14-16, D-53115 Bonn, Germany}

\author{Akram Hasani Zonoozi}
\affiliation{Helmholtz-Institut für Strahlen-und Kernphysik (HISKP), Universität Bonn, Nussallee 14-16, D-53115 Bonn, Germany}
\affiliation{Department of Physics, Institute for Advanced Studies in Basic Sciences (IASBS), PO Box 11365-9161, Zanjan, Iran}

\author{Pavel Kroupa}
\email{pkroupa@uni-bonn.de}
\affiliation{Helmholtz-Institut für Strahlen-und Kernphysik (HISKP), Universität Bonn, Nussallee 14-16, D-53115 Bonn, Germany}
\affiliation{Charles University in Prague, Faculty of Mathematics and Physics, Astronomical Institute, V Holešovičkách 2, CZ-180 00 Praha 8, Czech Republic}

\date{\today}

\begin{abstract}
   In the framework of Milgromian dynamics (MOND), galaxy clusters have been thought to have about a factor of two less baryonic mass than gravitational mass. One hypothesized source of this missing mass is undetected baryons. Extensive observations and studies indicate that the baryon content of galaxy clusters is primarily composed of the intracluster medium (ICM). In this work we re-evaluate the overall stellar mass in galaxy clusters taking into account recent work on the galaxy-wide stellar initial mass function of stars (gwIMF) needed to synthesise the metals observed in galaxies. Given their super-solar metallicities and short formation timescales, massive elliptical galaxies are inferred to have formed with highly top-heavy gwIMFs, which in turn leave behind a substantial mass in stellar remnants. The dependency of the gwIMF on the properties and evolution of a galaxy is well encapsulated by the integrated galaxy-wide initial mass function (IGIMF) theory, developed independently of MOND. 
   We utilize observational data at redshifts z <0.1 from the WIde-field Nearby Galaxy-cluster Survey (WINGS) and the Two Micron All Sky Survey (2MASS). Masses of galaxies and intracluster light (ICL) are calculated for 46 galaxy clusters using the IGIMF theory. The resulting masses in stars and in remnants are combined with previously derived ICM masses to estimate the total baryonic masses of the clusters. These baryonic masses are then compared to the MOND dynamical masses of the clusters, which are derived from hydrostatic equilibrium of the ICM based on earlier studies. As a complement, we include a comparison with several weak/strong lensing masses in the MOND framework. Our results show that the stellar masses of galaxies and the ICL increase substantially when applying the galaxy-wide mass-to-light ratios derived from the IGIMF theory. This leads to a significant rise in the estimated baryonic masses of galaxy clusters. In the sample of 46 galaxy clusters, the baryonic component on average accounts for $52^{+4}_{-3}\%$ of the MOND dynamical mass when considering only the ICM contribution. The baryonic mass in stars, remnants and the ICM accounts for at least $88^{+5+2}_{-4-1}\%$ of the MOND dynamical mass. The contribution by stellar remnants that arises from nucleosynthesis constraints thus significantly alleviates the missing mass problem in MOND. 
   Finally, we briefly discuss the compatibility of the IGIMF framework with the radial acceleration relation (RAR), and studies of MOND weak/strong lensing and related issues. A more comprehensive investigation will require future work that combines the IGIMF with self-consistent, spatially resolved formation, evolution, and resulting mass distribution models of galaxies.

\end{abstract}

\maketitle
\begin{center}
{\small Accepted for publication in \textit{Physical Review D}.}
\end{center}

\section{Introduction}
     The MOND theory, proposed as an alternative to dark matter, has achieved notable success in explaining galactic-scale dynamics as well as the dynamics of open star clusters \citep{kroupa2022asymmetrical,kroupa2024open}, while wide binary system are seen to be also successfully described \citep{hernandez2024critical, hernandez2024recent}.
     For general introductions and reviews of the MOND Theory, see, for example, \citet{sanders2002modified, 2006AIPC..822..253S, sanders2007modified, famaey2012modified, trippe2014missing, sanders2015historical, banik2022galactic, kroupa2023many, milgrom2014mond}.  
     
     However, when applied to galaxy clusters, MOND has long faced the issue of missing mass.     
     The masses of galaxy clusters can be estimated using several independent methods, including dynamical approaches based on the assumption of hydrostatic equilibrium of the intracluster gas or on measurements of the velocity dispersion of the cluster galaxy population, as well as gravitational lensing, which probes the projected mass distribution directly and does not rely on dynamical equilibrium. In recent years, several studies have also investigated strong- and weak-lensing masses of galaxy clusters within the MOND framework \citep{2020ApJTian,Famaey2025PhRvD,Mistele2025OJAp}.
     In the framework of Newtonian gravity, the mass inferred is typically larger by a factor of 5-10 compared to the mass of the observable matter, necessitating the introduction of dark matter to account for the gravitational influence in clusters. Today, the dark matter theory has been extensively studied from beyond galaxy cluster to galaxy scales, but many serious problems remain (see e.g., \citet{kroupa2023many}). For example, this dark matter ansatz is not supported by the lack of Chandrasekhar dynamical friction on the orbit of the Small and Large Magellanic Clouds and the Milky Way \citep{universe10030143, universe10100385, Hernandez2025Univ}.      
     But within the MOND framework, galaxy clusters still require an additional amount of unseen matter, roughly twice the observed baryonic mass, to explain their gravitational behavior. This problem has been extensively documented, see, for example, \citet{angus2008x} and \citet{brownstein2006galaxy}.

     Many efforts have been made to explain the missing mass in galaxy clusters within the MOND framework. Based on the assumption that galaxy clusters are in hydrostatic equilibrium, \citet{hodson2017generalizing} attempted to reduce the need for additional mass by extending MOND. \citet{lopez2022virial} reanalyzed the role of the virial theorem in determining MOND masses, which are fundamental to velocity dispersion-based mass measurements. In particular, they accounted for a correction to the surface pressure term in the virial theorem and adjusted the density models for galaxy clusters, which ultimately mitigated most of the missing mass.    
     \citet{angus200911} suggested that the problems in MOND including the missing mass of galaxy clusters could be explained by the introduction of 11 eV sterile neutrinos, which gives rise to the $\nu$HDM cosmological model \citep{katz2013galaxy,wittenburg2023,Samaras2025MNRAS}. 
     Alternatively, \citet{milgrom2015ultra} proposed cluster baryonic dark matter (CBDM) as a candidate for the missing mass, which may be related to the heating of the X-ray gas in galaxy clusters (\citet{Milgrom2008NewAR}), with ultra-diffuse galaxies (UDGs) offering a potential possibility to trace CBDM.
     
     The baryons in galaxy clusters are primarily composed of the intracluster medium (ICM), galaxies, and intracluster light (ICL or the diffuse light in a galaxy cluster). In previous studies, the observable baryonic mass was thought to be dominated by the ICM, with the stellar mass of galaxies and ICLs composing only about 15\% of the total baryonic mass, see, for example, \citet{lin2003near}. The ICM is assumed to be traced by X-ray gas, with temperatures of up to a few keV, which radiates through free-free emission (bremsstrahlung). Its mass can be estimated from X-band radiation observations. The stellar mass is typically estimated by using the mass-to-light ratios after luminosity measurements in corresponding photometric bands. In this work, rather than directly exploring candidates for the missing mass, we focus on re-calculating the stellar mass in galaxy clusters based on the IGIMF theory, which takes into account the variation of the stellar initial mass function across different star-forming environments, and which has been formulated and gauged using data entirely independently of MOND and galaxy clusters \citep{2024arXiv241007311K}.

     The IGIMF \citep{kroupa2003galactic} extends the stellar initial mass function (IMF) \citep{kroupa2001variation} to describe the empirical stellar mass distribution on galactic scales. The IGIMF theory, developed over the past two decades, has shown notable consistency with observational data across a variety of astrophysical contexts and methodologies. For example, \citet{Fontanot2017} demonstrated that the IGIMF yields an [$\alpha$/Fe]–stellar mass relation (with $\alpha$ referring to $\alpha$-elements such as O, Mg, and Si) that more accurately reflects observational trends. Likewise, \citet{Vincenzo2015} found that the theory successfully reproduces the observed abundance patterns in the Sagittarius dwarf galaxy. \citet{2024arXiv241007311K,Gunawardhana2011,Lee2009ApJ} show the  galaxy-wide initial mass function to vary systematically with star formation rate (SFR), as described by the IGIMF theory.   
     Additional studies reinforcing the agreement between the IGIMF framework and observations include \cite{Recchi2015,Weidner2013MN,weidner2013galaxy}. However, some studies have reported discrepancies between the IGIMF theory and observations. \citet{Andrews2013} argued for the independence of maximum stellar mass and cluster mass based on observations of the irregular galaxy NGC 4214. This conclusion was rebutted by \cite{Weidner2014}, who showed that the sample size and methodology in \cite{Andrews2013} were erroneous. 
     Similarly, \citet{Lacchin2020} found that the IGIMF theory could not reproduce the key chemical properties when modelling ultrafaint dwarf galaxies. In response, \citet{yan2020chemical} investigated one such galaxy, Bo\"otes I, and showed that adjusting two free parameters, namely the gas depletion timescale and the initial gas mass, which together determine the star formation history, allows the IGIMF framework to reproduce the observed chemical properties.

     Combining the IGIMF and the relationship between stellar mass and luminosity, the total stellar mass and luminosity of the whole galaxy can be determined, allowing for the calculation of the galaxy’s mass-to-light ratio. 
     The IGIMF theory correctly accounts for the observed variation of the stellar population in star-forming dwarf \citep{Lee2009ApJ} and massive \citep{Gunawardhana2011} disk galaxies. Concerning massive elliptical galaxies, their high (super-solar) metallicities require the galaxy-wide IMF (gwIMF) to have been top-heavy because of the need for a large population of massive stars for nucleosynthesis. This is naturally and correctly described by the IGIMF theory \citep{yan2021downsizing}.
     Unlike the mass-to-light ratios obtained from a direct application of the invariant canonical IMF, the IGIMF theory thus predicts that massive elliptical galaxies that follow the downsizing relation and experienced a top-heavy gwIMF during their early formation, formed a larger number of high-mass stars. Consequently, by redshift $z=0$, these galaxies contain a higher fraction of stellar remnants (see e.g., \citet{yan2017optimally,Zonoozi2025}). This has a particularly significant impact on the mass-to-light ratios of giant elliptical galaxies—especially the central dominant or brightest cluster galaxies (cD/BCGs)—as well as S0 galaxies. For the ICL it is also reasonable to use the mass-to-light ratio provided by the IGIMF to estimate its mass. The ICL consists mainly of old stellar populations, and is thought to originate from complex gravitational interactions within galaxy clusters, where stars are stripped from their parent galaxies. For a recent review of the origin and properties of the ICL, see, for example, \citet{contini2021origin}.

     To investigate the stellar mass in galaxy clusters, we compiled data from several major surveys and studies. We primarily utilized data from the WINGS project \citep{fasano2006wings,fasano2012morphology,d2014surface}, which provided morphologies and B,V-band luminosities of member galaxies in 21 galaxy clusters. We also supplemented this with data from the 2MASS project \citep{jarrett20002mass,lin2004k,lin2004kk}, which provided luminosity functions for 41 galaxy clusters and the luminosities of their BCGs in the K-band. One additional cluster not covered by WINGS or 2MASS is also used, it is the NGC 5044 group \citep{mendel2008anatomy}. In total, luminosity data for 63 galaxy clusters were compiled. These galaxy clusters were then screened to select a subsample that satisfies the assumptions of spherical symmetry and hydrostatic equilibrium. Clusters exhibiting clear evidence of major mergers were excluded from the analysis. To complete our dataset, we obtained the gas masses, Newtonian and MOND dynamical masses for the corresponding galaxy clusters from \citet{brownstein2006galaxy} (with partial data for the NGC 2563 and AWM4 groups from \citet{angus2008x}), allowing for a comprehensive analysis of the mass components in these systems. 

     We proceed to investigate how the variation of the galaxy-wide IMF previously extracted from a wide range of data (\citealt{2024arXiv241007311K,Jerabkova2025arXiv}) affects the mass in stars and in remnants of galaxy clusters. We find that the missing mass in MOND is largely accounted for as a result of the presence of a large amount of mass in remnants that arise due to the need to synthesise the super-solar metallicities of the massive early-type galaxies. These results are discussed in comparison to dynamical mass estimates from early-type galaxies that indicate these to be represented well by an invariant galaxy-wide IMF. The tension that arises in view of  the need to synthesise the elements and to understand the stellar-dynamical constraints is discussed pointing to further work that is needed to study the spatial dispersion of remnants within and around early-type galaxies.

     The structure of this paper is organized as follows:
     In Sect. \ref{BasicTheo}, we present the theoretical foundations, including gas mass measurements, the dynamical mass measurements in both Newtonian and MOND regimes, and the principles of the IMF and the IGIMF theory.
     Section \ref{DataGC} covers the datasets used and their processing, including galaxy population data from WINGS, luminosity function parameters from 2MASS, and data of NGC 5044. The ICL fraction is also introduced.
     In Sect. \ref{SteMaBoost}, we derive the mass-to-light ratios of galaxies under the assumption that the galaxy-wide IMF is either the invariant canonical IMF or that it is given by the IGIMF, with separate analyses across different bands.
     Section \ref{Result} presents a comparison of results across various samples, featuring comparative plots of different mass components. A complete summary of the sample is provided, with a final comparison plot of total baryonic mass and MOND dynamical mass. We also compare our results with a recent study on the missing mass in galaxy clusters under the MOND framework (\citet{Kelleher:2024oip}).
     Finally, in Sect. \ref{clarify} we discuss the potential tension between the IGIMF theory and dynamical or lensing observations of galaxies, and we provide a brief summary of studies that are either consistent with our results or in conflict with them.

     All error bars are within the $1\sigma$ range, corresponding to a 68\% confidence level, unless otherwise noted. The Hubble constant adopted throughout is $H_{0} = 70 $km/s/Mpc such that $h \equiv H_{0}/(100$ km/s/Mpc). The absolute magnitude system used from the dataset is the Vega system.
     
\section{Basic theory}
\label{BasicTheo}
     In this section, we provide an overview of the concept of dynamical mass in galaxy clusters, the principles underlying gas mass measurements, and the IGIMF theory.
     
\subsection{Gas mass}
     The X-ray gas, which dominates the gas distribution, is considered a tracer for the ICM, allowing the ICM (total gas) mass to be estimated from the observed X-ray gas mass. A commonly used model for approximating the spherical distribution of the gas density is the $\beta$-model \citep{cavaliere1976x,cavaliere1978distribution},
     \begin{equation}\label{Beta}
     \begin{aligned} 
     \rho(r) = \rho_{0}\left[1 + \left(\frac{r}{r_{\rm c}}\right)^{2}\right]^{-\frac{3\beta}{2}}.
     \end{aligned}
     \end{equation}
     Here $\rho(r)$ is the mass density of the gas at a distance $r$ from the cluster center; $\rho_{0}$ is the core density; $r_{\rm c}$ is the core radius; $\beta$ is a fitting parameter. The parameters $\rho_{0}$, $r_{\rm c}$, and $\beta$ are typically determined by fitting the surface brightness profile of the X-ray gas, from which the gas density distribution can be derived. The total gas mass within a radius $r$ is then obtained by integrating the density over the volume \citep{sarazin1986x,brownstein2006galaxy}. For the X-ray gas with $r_{\rm c} \ll r$, and $\beta < 1$, the total gas mass can be approximated as
     \begin{equation}\label{Beta_mass}
     \begin{aligned} 
     M(r) \approx \frac{4\pi \rho_{0} r^{3}_{\rm c}}{3(1-\beta)}\left(\frac{r}{r_{\rm c}}\right)^{3(1-\beta)} .
     \end{aligned}
     \end{equation}
     In this study, we adopt the gas mass within the virial radius, assuming the virial radius $r_{\rm vir} \approx$ $r_{200}$ from the Newtonian gravity framework. $r_{200}$ represents the characteristic radius in the Newtonian gravity framework within which the average mass density is 200 times the critical density of the universe. The gas masses for all the galaxy clusters in our sample are taken from \citet{brownstein2006galaxy} and \citet{angus2008x}. 
     
\subsection{Dynamical mass from hydrostatic equilibrium}
\label{dynamical_Theory}
     The temperature of the X-ray-emitting gas in galaxy clusters typically reaches up to a few keV. By assuming that the gas is in isothermal hydrostatic equilibrium, it becomes possible to estimate the gravitational potential of the galaxy cluster, and thus derive its total mass. Under this assumption, the gas model must satisfy the following equation \citep{fabricant1980x}
     \begin{equation}\label{Hydro}
     \begin{aligned} 
     \frac{\mathrm{d}P(r)}{\mathrm{d}r} = -\rho(r)\frac{\mathrm{d}\Phi(r)}{\mathrm{d}r},
     \end{aligned}
     \end{equation}
     where $P(r)$ is the pressure at radius $r$; $\Phi(r)$ is the gravitational potential at radius $r$. 
     
     In the Newtonian framework, the total dynamical mass of the galaxy cluster within radius $r$ is
     \begin{equation}\label{NewDynMass}
     \begin{aligned} 
     M_{\rm N,dyn}(r) = -\frac{k_{\rm B}T(r)r}{\mu m_{\rm p}G}\left[\frac{\mathrm{d}\ln(\rho(r))}{\mathrm{d}\ln r} + \frac{\mathrm{d}\ln(T(r))}{\mathrm{d}\ln r}     \right].
     \end{aligned}
     \end{equation}
     Here $k_{\rm B}$ is the Boltzmann constant; $T(r)$ is the temperature distribution of the gas; $\mu \approx 0.609$ \citep{brownstein2006galaxy} is the mean atomic weight (as the ratio of the mean atomic mass to the proton mass); $m_{\rm p}$ is the mass of proton; $G$ is the gravitational constant.
     Once the $\beta$-model in Eq.~(\ref{Beta}) and the isothermal assumption are applied, the derivative term of the temperature $T$ vanishes and we get the approximate form
     \begin{equation}\label{NewDynMass2}
     \begin{aligned} 
     M_{\rm N,dyn}(r) = \frac{3\beta k_{\rm B}T(r)}{\mu m_{\rm p} G} \left(\frac{r^{3}}{r^{2} + r^{2}_{\rm c}}\right).
     \end{aligned}
     \end{equation}

     In the MOND framework the relationship between the physical gravitational acceleration $g(r) \equiv -\mathrm{d}\Phi(r)/\mathrm{d}r$ and the Newtonian acceleration $g_{\rm N}$ is
     \begin{equation}\label{MondAcc}
     \begin{aligned} 
     \mu_{\rm MOND}\left(\frac{g}{a_{0}}\right) g= g_{\rm N}.
     \end{aligned}
     \end{equation}
     Here $a_{0} \approx 1.2 \times 10^{-10}$ $\mathrm{ms^{-2}}$, is the Milgromian constant; the interpolating function $\mu_{\rm MOND}$ is a function of $g$ and has multiple empirical formulations. In \citet{brownstein2006galaxy} the form
     \begin{equation}\label{MondMu}
     \begin{aligned} 
     \mu_{\rm MOND}(x) = \frac{x}{\sqrt{1+x^{2}}},
     \end{aligned}
     \end{equation}
     is used, which leads to the expression of the MOND dynamical mass within $r$,
     \begin{equation}\label{MondMass1}
     \begin{aligned} 
     M_{\rm M,dyn}(r) = \frac{M_{\rm N,dyn}(r)}{\sqrt{1 + \left(\frac{a_{0}r^{2}}{GM_{\rm M,dyn}(r)}\right)^{2}}},
     \end{aligned}
     \end{equation}
     or, equivalently,
     \begin{equation}\label{MondMass2}
     \begin{aligned} 
     M_{\rm M,dyn}(r) = \sqrt{M^{2}_{\rm N,dyn}(r) - \frac{a^{2}_{0}r^{4}}{G^{2}}}.
     \end{aligned}
     \end{equation}
     In this work, the Newtonian and MOND dynamical masses of all samples are also taken from \citet{brownstein2006galaxy} and \citet{angus2008x}. Note that \citet{brownstein2006galaxy} employed convergent MOND masses measured within $r_{\rm out}$, defined as the radius where the density drops to 250 times the mean cosmological baryon density. In this work, we approximate this mass as the mass within $r_{\rm 200}$.

\subsection{The mass-to-light ratio of a stellar population from the IGIMF theory}
\label{IGIMF}
     To convert the luminosity of a galaxy into its stellar mass, we need to obtain the galaxy's stellar mass-to-light ($M/L$) ratio. Note that the stellar $M/L$ ratio used in this work is defined as the total mass in stars and their remnants ($M$) divided by their total luminosity ($L$) in a given photometric band. Accordingly, the stellar masses in this study always include the contribution from stellar remnants. We used the stellar population synthesis technique to simulate galaxies with different properties and calculate their stellar $M/L$ ratios at different bands. To do this one requires assumptions about the underlying gwIMF, the star formation history of the galaxy, and its metallicity.  The gwIMF is adopted to be invariant in most existing stellar population synthesis (SPS) models. Here we compare the $M/L$ ratios of modeled galaxies being calculated with the invariant canonical IMF as the gwIMF \citep{kroupa2001variation} with those being calculated with the star-forming-gas-density- and metallicity-dependent integrated galaxy-wide IMF (IGIMF).  

     For the invariant canonical IMF we use the canonical 2-segment power-law function as follows
     \begin{eqnarray}
     \xi_\star(m)=k_\star\left\{
     \begin{array}{ll}
     \,2 m^{ -\alpha_1},      & 0.08M_\odot\leq m < 0.5 M_\odot,  \\
     \,m^{ -\alpha_2},      & 0.5M_\odot\leq m < 150 M_\odot,\\
     \end{array}
     \right. \label{MF1}
     \end{eqnarray}
     where $k_\star$ is the normalization constant.
     The number of stars that form in the mass interval $m$ to $m+\mathrm{d}m$ is $\mathrm{d}N_\star=\xi_\star(m)\mathrm{d}m$. The minimum stellar mass is set to 0.08 $M_{\odot}$, corresponding to the hydrogen-burning limit (see e.g., \citet{Thies2015ApJ,2024arXiv241007311K}). The less-massive brown dwarfs contribute a negligible amount of mass to the known stellar population. The maximum stellar mass in an embedded cluster is assumed to be 150 $M_{\odot}$ (see also \citet{yan2017optimally}). The canonical stellar IMF has $\alpha_1=1.3$, and $\alpha_2=2.3$ (see \citet{2024arXiv241007311K}). 

     Most stars form within embedded clusters \citep{kroupa1995inverse, kroupa1995dynamical, lada2003embedded, Kroupa05, megeath2015spitzer, dinnbier2022majority}. These clusters, regardless of whether they have dissolved, contribute to the stellar population of the galaxy. Accordingly, the gwIMF is derived by summing the individual stellar IMFs of all embedded clusters that formed during a specific time interval, $\delta t$.
     If the stellar IMF is an invariant canonical form (Eq.~\ref{MF1}), then this sum leads to the same form. But if the stellar IMF varies with the physical conditions of the molecular cloud clumps that spawn embedded stellar clusters, then the summed (i.e., integrated) galaxy-wide IMF (i.e., IGIMF) will differ in form from the stellar IMF as given by  Eq.~(\ref{MF1}).
     
     The stellar IMF, which depends on the metallicity and density of the star-forming gas clumps, can be characterized as a multi-power-law function, as formulated in \cite{kroupa2001variation, kroupa2002initial}, \cite{yan2020chemical, yan2021downsizing}, and \citet{Haslbauer2024}. Assuming that the upper limit for stellar mass, $m_{\rm max}$, is higher than $1.0M_\odot$, the stellar IMF can be expressed as follows:
     \begin{eqnarray}
     \xi_\star(m)=k_\star\left\{
      \begin{array}{ll}
     \,2 m^{ -\alpha_1},      & 0.08M_\odot\leq m < 0.5 M_\odot,  \\
     \,m^{ -\alpha_2},      & 0.5M_\odot\leq m < 1.0M_\odot,\\
     \,m^{ -\alpha_3},      & 1.0M_\odot\leq m < m_{\rm max}.\\
       \end{array}
        \right. \label{MF2}
     \end{eqnarray}
     In general, $\alpha_1$ and $\alpha_2$ are functions of metallicity, while $\alpha_3$ is a function of both metallicity and the initial mass of the embedded cluster and $m_{\rm max}=m_{\rm max}(M_{\rm ecl})$ is the most massive star that forms in the embedded cluster with a mass in stars of $M_{\rm ecl}$ (\citet{Yan2023}; see also \citet{jevrabkova2018impact}, \citet{Zonoozi2025} and \citet{Gjergo2026RAA} for a detailed explanation).

     In order to calculate the gwIMF of all the stars formed in the galaxy during the time interval $\delta t$, we need to sum the IMFs of all embedded clusters. With $\mathrm{d}N_{\rm ecl}=\xi_{\rm ecl}(M_{\rm ecl})\mathrm{d}M_{\rm ecl}$ being the number of embedded clusters within the mass interval of $M_{\rm ecl}$ and $M_{\rm ecl}+\mathrm{d}M_{\rm ecl}$, $\xi_{\rm ecl}(M_{\rm ecl})$ is the mass function of these embedded clusters, known as the Embedded Cluster Mass Function (ECMF). We assume it to be a single power-law distribution with a slope $\beta_{*}$ that depends on the SFR \citep{weidner2013galaxy, zhang2018stellar, recchi2009chemical, yan2017optimally},
     \begin{equation} \label{MF3}
     \xi_{\rm ecl}(M_{\rm ecl}) =
     \begin{cases}
     0, & M_{\rm ecl} < M_{\rm ecl,min},\\[2mm]
     k_{\rm ecl} M_{\rm ecl}^{-\beta_*}, & M_{\rm ecl,min} \le M_{\rm ecl} < M_{\rm ecl,max}(SFR),\\[1mm]
     0, & M_{\rm ecl} \ge M_{\rm ecl,max}(SFR),
     \end{cases}
     \end{equation}
     where
     \begin{equation}
     \beta_{*}=-0.106 \log_{10} \psi(t) + 2, 
     \label{beta}\\
     \end{equation}
     where $\psi(t)$ represents the SFR in units of $M_\odot/\rm{yr}$. Specifically, Eq.~(\ref{MF3}) is taken from the model of \citet{yan2017optimally}, where the single-slope power-law assumption is motivated by observational constraints \citep{lada2003embedded,Weidner2004MNRAS}. Eq.~(\ref{beta}) follows the assumption of \citet{weidner2013galaxy}, introduced to reproduce the constraints of \citet{Gunawardhana2011} on the gwIMF of massive star forming disk galaxies. 
     The minimum stellar mass of embedded clusters is set at $M_{\rm ecl,min}= 5 M_{\odot}$ , based on the smallest observed star-forming molecular cloud clumps \citep{kroupa2003origin, kirk2012variations, Joncour2018}. The maximum mass of embedded star clusters, denoted as $M_{\rm ecl,max}$, is dependent on the SFR.  Both  $M_{\rm ecl,max}$ and normalization constant, $k_{\rm ecl}$, can be determined by simultaneously solving the following equations
     \begin{eqnarray}
     M_{\rm tot,\delta t}=\int_{M_{\rm ecl,min}}^{M_{\rm ecl,max}}\xi_{\rm ecl}(M)  M \mathrm{d}M=\psi(t)\delta t, \nonumber\\
     1=\int_{M_{\rm ecl,max}} ^{10^9 M_\odot} \xi_{\rm ecl}(M) \mathrm{d}M,\ \ \ \ \ \ \ \ \ \ \ \ \ \ \ \ \ \ \ \ \ \ \ \ \label{Mtot10}
     \end{eqnarray}
     where $M_{\rm tot,\delta t}$ is the total stellar mass that forms in a galaxy during a single star formation epoch lasting $\delta t= 10$ Myr, which corresponds to the typical lifetime of molecular clouds (see \citet{kroupa2013, schulz2015mass}, \citet{yan2017optimally} and \citet{2024arXiv241007311K} for a discussion and additional references). 

     The first part of Eq.~(\ref{Mtot10}) represents the total mass formed in 10 Myr, denoted as $M_{\rm tot,\delta t}$. The second part indicates that only one cluster is formed within the range of the maximum embedded cluster mass ($M_{\rm ecl,max}$) and the upper cluster mass limit (assumed here to be $10^9 M_{\odot}$, but the results are not sensitive to $M_{\rm ecl,max}>10^9 M_{\odot}$). 

     With the stellar IMF and ECMF established, we can construct the stellar IMF for the entire galaxy by summing the IMFs of all star clusters that form during the time interval $\delta t$.

     Using the \texttt{SPS-VarIMF} code developed by \citet{Zonoozi2025},  we calculated the stellar $M/L$ ratios of galaxies in different bands for the invariant canonical IMF as the gwIMF and the IGIMF theory assuming a constant and a varying gwIMF with time, a SFR and an assumed metalicity. Since galaxy clusters include both star-forming and quiescent galaxies, we calculate the $M/L$ ratios for two different types of galaxies: spiral galaxies with ongoing star formation which is assumed to be constant over the lifetime of galaxies (i.e., 12 Gyr; \citet{Kroupa2020, Haslbauer2023}), and elliptical galaxies which are quiescent in terms of star formation. The formation time scale of elliptical galaxies is adopted based on the downsizing relation \citep{thomas2005epochs, mcdermid2015atlas, yan2021downsizing} as follows
     \begin{equation}
     \tau_{\rm SF}[\rm Gyr]=49 (M/M_{\odot})^{-0.14}, \label{downsizing}\\
     \end{equation}
     which means more massive galaxies tend to form their stars earlier and over shorter time scales than less massive galaxies. The star formation is assumed to be constant during this period.  The formation timescales of massive elliptical galaxies are discussed in \citet{2025NuPhBGjergo}.  We emphasize that BCGs are also assumed to follow the downsizing relation. Observations have revealed highly overdense protocluster cores at high redshifts (see e.g., \citet{Ishigaki2016ApJ,Jiang2018NatAs,Miller2018Natur}). Since protocluster cores are considered the birth sites of BCGs (see e.g., \citet{Ito2019ApJ}), this provides support for the scenario in which BCGs formed rapidly at high redshift (see also \citet{Rennehan2020MNRAS}).

     The adopted assumptions for the star formation histories of spiral and elliptical galaxies are consistent with observations. Spiral galaxies, as well as low-mass galaxies, are observed to sustain nearly constant SFRs \citep{Kroupa2020,Haslbauer2023,Yang2024ApJ}. In contrast, massive elliptical galaxies exhibit very high SFRs at high redshift \citep{mcdermid2015atlas,Nguyen2020ApJ}. The most massive elliptical galaxy had initial SFRs  exceeding 1000 $M_{\odot}/\mathrm{yr}$, a value that is consistent with observational evidence \citep{Nguyen2020ApJ}.     
     
\begin{figure}
     \centering
     \includegraphics[width=0.47\textwidth]{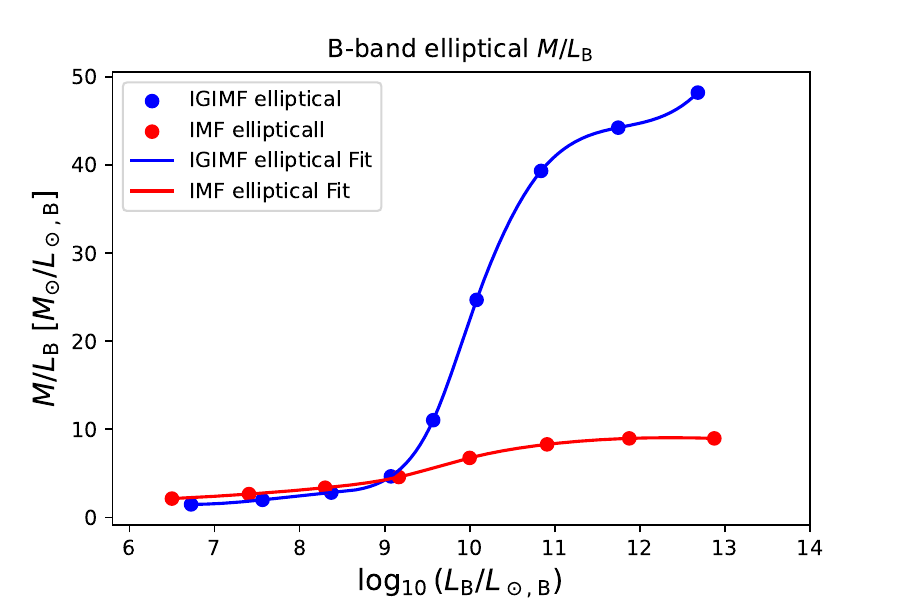}
     \includegraphics[width=0.47\textwidth]{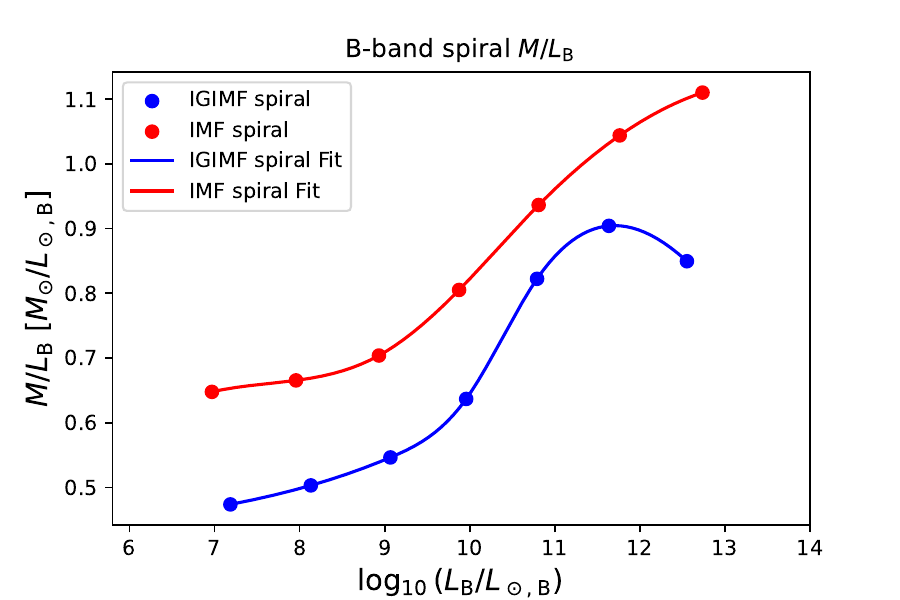}
     \caption{The relation between $M/L$ ratios and luminosities in the B-band at redshift $z = 0$. Upper plot: The relation for elliptical galaxies. The blue dots are the $M/L$ values in the IGIMF theory, the blue line is the result of cubic spline interpolation to the blue dots. The red dots are the $M/L$ values in the canonical IMF Theory, the red line is the result of cubic spline interpolation to the red dots. Lower plot: the relation for spiral galaxies, otherwise as above.}
     \label{Bband}
     \end{figure}

     Interestingly, the downsizing relation follows naturally from the collapse of proto-galactic gas clouds in MOND \citep{Eappen2022, Eappen2024}. The downsizing relation implies that massive elliptical galaxies include older stars and more remnants. The larger mass in stellar remnants and low-mass stars in elliptical galaxies results from the large SFRs and high metallicities, respectively (see \citet{yan2021downsizing}; \citet{2024arXiv241007311K}). This results in a large $M/L$ ratio for massive elliptical galaxies. For spiral galaxies, the $M/L$ ratios predicted by the IGIMF do not differ significantly from those obtained with a canonical IMF. However, in the case of low-mass spiral galaxies with small SFRs, the IGIMF predicts a different evolutionary history. Specifically, such low SFRs ($<10^{-3}$ $\mathrm{M_{\odot}/yr}$; consistent with observation; \cite{2024arXiv241007311K,Jerabkova2025arXiv} for review) imply that no stars more massive than $4 M_{\odot}$ can form \citep{yan2017optimally}. Consequently, these galaxies do not experience early evolution driven by massive stars, and their luminosities are instead modified at later stages through processes such as the red-giant branch evolution. A more detailed description of such evolutionary pathways can be found in \citet{Zonoozi2025} (see also Fig.~\ref{Bband}).

     Metallicity is a key factor influencing both the luminosities and masses of galaxies. In general, stars with low metallicity tend to appear bluer and emit more ultraviolet light compared to their metal-rich counterparts. As the metallicity increases, the opacity in stellar atmospheres rises, which lowers the effective temperature of stars and shifts the integrated galaxy colors toward the red. Consequently, the stellar mass–metallicity relation implies that massive spiral galaxies, having experienced more extensive star formation in the past, are expected to be redder. In elliptical galaxies, the rapid quenching of star formation removes the population of massive, luminous blue stars, further enhancing their red colors relative to systems with sustained star formation activity.     
     In our models, galaxies of the same stellar mass are assigned a constant metallicity, $Z$. The metallicities of galaxies with different stellar masses follow the relation of \citet{gallazzi2005ages}. This relation yields values ranging from $Z=0.03$ at stellar mass $M = 10^{13}M_{\odot}$ to $Z = 0.0002$ at $M = 10^{5}M_{\odot}$ \citep{Zonoozi2025}.     
     Under these metallicity assignments, we compute the $M/L$ ratios of both spiral and elliptical galaxies assuming either the invariant canonical IMF or the IGIMF prescription.  

     In Fig.~\ref{BK_evo} and \ref{Vr_evo}, we also present the $M/L$ ratios of elliptical galaxies
     assuming galactic chemical enrichment in the IGIMF framework, for comparison. The uncertainties are estimated from the error bounds of the cubic-spline interpolation \citep{hall1976optimal}. In this case, the initial metallicity of all galaxies is set to $Z=0.0002$, which subsequently enrich to the adopted constant metallicity value by $z=0$.       
     As shown, the $M/L$ ratios in the B-band differ only slightly from those obtained under the constant-metallicity assumption. In contrast, the r- and V-bands $M/L$ ratios increase for galaxies with luminosities $> 10^{11.5} L_{\rm \odot, r}$ and $L_{\rm \odot, V}$, respectively, while the K-band $M/L$ ratio becomes significantly higher already at $L > 10^{10.5} L_{\rm \odot, K}$.      
     This occurs because the constant-metallicity galaxies have a higher metallicity than the average metallicity of the evolving case, resulting in brighter luminosities at longer wavelengths. At lower metallicity, the gwIMF is more top-heavy and remnant masses are also higher at low metallicity. Thus, low-metallicity cases tend to have higher $M/L$ ratios. The effect is particularly pronounced for more metal-rich, high-luminosity galaxies. Consequently, in the K-band (bottom panel of Fig.~\ref{BK_evo}), massive elliptical galaxies exhibit lower $M/L$ ratios under the constant metallicity assumption (see \citet{Zonoozi2025MNRASBTFR}).
     In this work, we adopt a conservative approach by employing the constant-metallicity model in the calculations, which provides a lower limit on the $M/L$ ratios.

     The top panel of Fig.~\ref{Bband} compares the calculated B-band $M/L$ ratio of elliptical galaxies assuming a canonical invariant IMF and the IGIMF. It follows that the $M/L$ ratios of massive elliptical galaxies are about 4-6 times larger in the context of the IGIMF compared to the canonical IMF (see also \citet{yan2021downsizing}). The lower panel of Fig.~\ref{Bband} shows the comparison of $M/L$ ratios for spiral galaxies. It can be seen that the values derived with the canonical IMF and the IGIMF are broadly consistent, with differences of only about 10\%-20\%.

\section{Galaxy cluster data}
\label{DataGC}
     In this section we introduce the used data, and the processing of the data. 
     
\subsection{Data from WINGS}
     We identified 21 galaxy clusters in the WINGS dataset \citep{fasano2006wings} that match the cluster list from \citet{brownstein2006galaxy}. 
     Clusters that are clearly undergoing mergers or significantly deviate from spherical symmetry were excluded. As a result, a final sample of 12 galaxy clusters was selected from the WINGS survey.   
     The B- and V-band apparent magnitudes of member galaxies in each cluster were obtained from \citet{d2014surface}, while the morphological classifications were taken from \citet{fasano2012morphology}. Since the $M/L$ ratio of a galaxy is strongly correlated with its morphological type (see Fig.~\ref{Bband}), we categorized the galaxies into two groups: elliptical and S0 galaxies use the $M/L$ ratio typical of elliptical galaxies to estimate their masses, while the remaining galaxies follow the $M/L$ ratio of spiral galaxies. In \citet{fasano2012morphology}, galaxy morphology is determined by a morphology parameter, where values $\leq 0$ correspond to S0 and elliptical galaxies, which are assigned the $M/L$ ratio of elliptical galaxies, and values $> 0$ correspond to spiral galaxies, which are assigned the $M/L$ ratio of spiral galaxies. Throughout the remainder of this paper, lenticular (S0), elliptical (E), dwarf S0 (dS0), and dwarf elliptical (dE) galaxies are collectively referred to as elliptical (or early-type) galaxies—including their associated $M/L$ ratios—unless otherwise specified. Similarly, all other galaxy types (e.g., spiral, irregular) are collectively referred to as spiral (or late-type) galaxies, along with their corresponding $M/L$ ratios.

\begin{figure}
     \centering
     \includegraphics[width=0.45\textwidth]{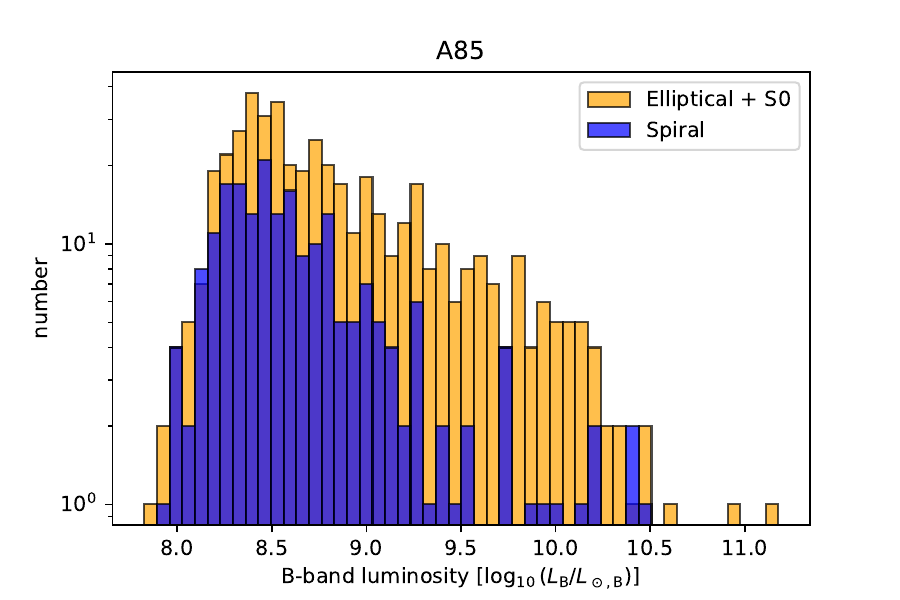}
     \caption{The B-band luminosity distribution of member galaxies in the galaxy cluster Abell 85. The orange histogram represents the number distribution of elliptical (E+dE) and S0 (S0+dS0) galaxies, while the blue histogram shows the number distribution of all other types of galaxies. Apparent magnitudes are from \citet{d2014surface}, while the morphologies are from \citet{fasano2012morphology}.}
     \label{a0085}
     \end{figure}

     For a galaxy with an apparent magnitude $m_{\rm B}$ in the B-band, its luminosity is approximated as:
     \begin{equation}\label{MagLumi}
     \begin{aligned} 
     L_{\rm B} = 10^{0.4(M_{\rm \odot,B} - M_{\rm B})}L_{\rm \odot,B},
     \end{aligned}
     \end{equation}
     where
     \begin{equation}\label{magMag}
     \begin{aligned} 
     M_{\rm B} = m_{\rm B} + 5\log_{10}\left(\frac{d_{0}}{d}\right).
     \end{aligned}
     \end{equation}
     Here $L_{\rm B}$ is the absolute B-band luminosity of the galaxy; $M_{\rm B}$ is the B-band absolute magnitude of the galaxy; $d_{0} = 10$ $\mathrm{kpc}$ is the reference distance; $d$ is the distance to the Earth in kpc, estimated from the redshift value of the corresponding galaxy cluster;  $M_{\rm \odot,B}$ is the B-band absolute magnitude of the Sun; $L_{\rm \odot,B}$ is the B-band luminosity of the Sun. The B-band luminosity distribution of member galaxies in the galaxy cluster Abell 85 is shown in Fig.~\ref{a0085} as an example. 

     Since the B-band, being at shorter wavelengths, can be affected by dust absorption, it is necessary to cross-check with infrared data. For this purpose, we used the K-band photometry from \citet{d2014surface} to examine a subset of cluster galaxies for which stellar masses had been estimated from the B- and V-bands. The K-band data are, however, much more limited than those in the B- and V-bands (about 1800 galaxies in the K band, compared to more than 10000 in the B and V bands across our initial 21 WINGS clusters), and were therefore not employed in the subsequent cluster stellar mass calculations. The comparisons of IGIMF and canonical IMF cases are shown in Fig.~\ref{BVK_IGIMF} and Fig.~\ref{BVK_IMF}, respectively. Taking the K-band stellar mass as a reference, we find that the masses derived from the B- and V-bands tend to be systematically lower, particularly for galaxies with $M < 10^{10} M_{\odot}$. However, the numerous low-mass galaxies contribute less to the total stellar mass budget than the massive galaxies, particularly in the IGIMF framework. When summing the stellar masses of all galaxies across different bands, we indeed find that the total mass derived from the $K$-band, amounting to about 80\% of that from the $V$-band, is comparable to the $V$-band estimate.
     
     In addition, we constructed a comparison between the observed colors (B-V and V-K) of elliptical galaxies in the Abell 119 cluster from the WINGS and the predictions from both the IGIMF and the canonical IMF models (see Fig.~\ref{color_compare}). It can be seen that the colors of galaxies predicted by both the IGIMF and the canonical IMF lie close to the core of the observed distribution. This indicates that our stellar population models reproduce the central values of the observations, while the larger scatter seen in the real data is likely due to factors such as dust absorption and detailed differences in star formation histories.

     Finally, it should be noted that the maximum radius $r_{\rm out}$ used to calculate the MOND dynamical masses, as mentioned in Sect.~\ref{dynamical_Theory}, is larger than the spatial extent of the member galaxies in the WINGS cluster catalogues. As a result, the stellar masses of the clusters may be underestimated.

\subsection{Data from 2MASS}
\label{2MASSData}
     We identified 41 common galaxy clusters in \citet{lin2004k} compared to those in \citet{brownstein2006galaxy} and \citet{angus2008x}. Following the screening, 33 galaxy clusters remain that satisfy the conditions of spherical symmetry and hydrostatic equilibrium. For these 33 clusters, we used the K-band data provided in \citet{lin2004k} to construct the luminosity function using the Schechter function \citep{schechter1976analytic}. It is noted that the luminosity function provides the best fit when excluding the contribution from the BCG \citep{sarazin1986x}. Therefore, we read the K-band extrapolated luminosity of the BCGs for each cluster from \citet{lin2004kk} and manually included BCG masses in the final mass estimates. Also note that some galaxy clusters may host more than one high-luminosity or cD galaxy, so this method may in fact underestimate the total cluster mass in such cases.
     
     The Schechter function in the K-band is given by
     \begin{equation}\label{Schehter}
     \begin{aligned} 
     \Phi(M_{\rm K}) = \frac{\Phi^{*}\log10}{2.5}  10^{0.4(\alpha+1)(M^{*}_{\rm K} - M_{\rm K})} \exp\left[-10^{0.4(M^{*}_{\rm K} - M_{\rm K})}\right],
     \end{aligned}
     \end{equation}
     where $\Phi^{*}$, $\alpha$ and $M^{*}_{\rm K}$ are fitting parameters; $\Phi(M_{\rm K})$ is the density distribution of galaxies in absolute magnitude space. $\Phi(M_{\rm K})\mathrm{d}M_{\rm K}$ denotes the number of galaxies in the magnitude interval $[M_{\rm K},M_{\rm K}+\mathrm{d}M_{\rm K}]$. In \citet{lin2004k}, the $\alpha$ is fixed at -1.1 for all clusters. In our analysis, $\Phi^{*}$ is renormalized so that the total number of galaxies matches the cluster population in \citet{lin2004k}. All galaxy counts derived from the luminosity function are rounded (rounding to the nearest integer), as will be the case for subsequent analyses.
     
     Since we do not have data on galaxy morphology for these 33 clusters, we estimate the ratio of elliptical and spiral galaxies based on the morphological proportions observed in neighboring galaxy
     clusters. According to Fig.~8 of \citet{fasano2015morphological}, within $r_{200}$, elliptical and S0 galaxies account for approximately 60\% of the total observable galaxy population. 
     We acknowledge that this method introduces uncertainty into the mass calculation. However, due to the significant increase in mass associated with the BCG in the IGIMF theory, this uncertainty is unlikely to affect the overall qualitative conclusions.

\subsection{Data for the NGC 5044}
\label{Sec_For_NGC}
     As noted by \citet{angus2008x}, the missing mass problem in galaxy clusters under the MOND framework is most pronounced in low-mass clusters. To address this, we have included additional data from a low-mass galaxy group in our study.   

\begin{figure}
     \centering
     \includegraphics[width=0.45\textwidth]{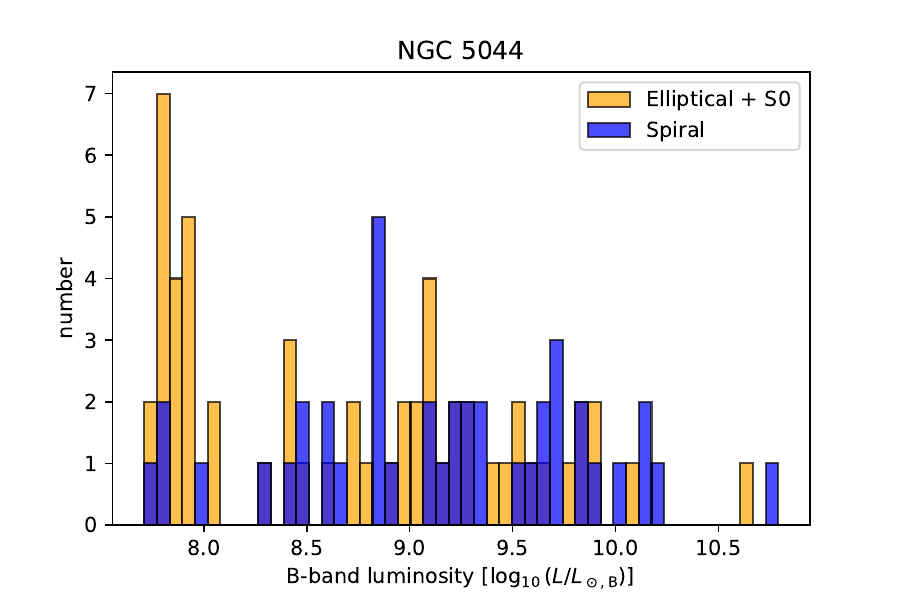}
     \caption{The B-band luminosity distribution of member galaxies in the galaxy group NGC 5044. The orange histogram shows the number distribution of elliptical and S0 galaxies, the blue histogram shows the number distribution of all other galaxy types. Data are from \citet{mendel2008anatomy}.}
     \label{5044}
     \end{figure}

     From \citet{mendel2008anatomy}, we obtained the galaxy dataset for NGC 5044, including B-band apparent magnitudes and galaxy morphology information. For galaxies without morphology data, we classified them as spiral galaxies. The B-band luminosity distribution of the member galaxies in the NGC 5044 galaxy group is shown in Fig.~\ref{5044}.
     
\subsection{ICL luminosity fraction}
\label{Sec_ICL_toy}
     The ICL constitutes a significant fraction of the stellar luminosity in galaxy clusters. 
     Since both the effective radius of galaxies in \citet{d2014surface} and the fixed metric radius for photometry in \citet{lin2004kk} are small, we find that the measured radius of the BCGs in these datasets do not significantly overlap with the ICL. Therefore, the stellar luminosity contributed by the ICL must be calculated separately.   

     The origin, identification and mass estimation of the ICL are still under investigation (see, e.g., \citet{pierini2008diffuse, sand2011intracluster, toledo2011diffuse, cui2014characterizing, harris2017quantifying, sampaio2021diffuse, Furnell2021MNRAS,contreras2024characterising}). Galaxy stripping has long been considered a leading hypothesis for the origin of the ICL. However, recent observations provide evidence pointing to an alternative picture. For instance, \citet{Joo2023} showed that clusters at $z > 1$ already contain a substantial amount of ICL, supporting scenarios in which most of the ICL is formed concurrently with the growth of the BCG, or through the accretion of pre-processed intracluster stars, rather than being predominantly produced by gradual stripping processes. A similar conclusion was reached by \citet{Werner2023MNRAS}, who reported the detection of significant ICL at $z \approx 2$. The $M/L$ ratio of the ICL under a canonical IMF may be underestimated (see \citet{tang2018investigation}).

     The mass contribution of the ICL is commonly characterized by the ICL fraction, which can be defined either as a stellar mass fraction, $M_{\rm ICL}/M_{\rm star}$, or as a luminosity fraction, $L_{\rm ICL}/L_{\rm star}$, where $M_{\rm star}$ and $L_{\rm star}$ denote the total stellar mass and luminosity of the galaxy cluster, respectively. Different studies also adopt different radial apertures, such as $r_{500}$ (defined at 500 times the critical density of the Universe, analogous to $r_{200}$) or $r_{200}$. In this work, we estimate the ICL mass using the ICL luminosity fraction measured within $r_{200}$. To this end, we combine results from recent studies to infer a representative ICL fraction, which is then applied to all galaxy clusters without observational data. 

     From the observational perspective, several studies of individual galaxy clusters report ICL fractions of about $20\%(+10\%, -10\%)$ (see, e.g., \citet{montes2021buildup,Oliveira2022MNRAS,Spavone2024A&A,Teja2025A&A}).        \citet{sand2011intracluster} estimated the ICL mass fraction based on supernova observations in clusters, finding an average value of $16\%(+13\%, -9\%)$. Assuming that the ICL is entirely composed of old stellar populations, they derived an upper limit of about $47\%$. \citet{ragusa2021vegas} compiled results from several clusters observed in different bands across multiple studies, reporting an ICL luminosity fraction of $\approx 20\%(+10\%, -10\%)$. More recently, \citet{ragusa2023does}, using data from the VST Early-type Galaxy Survey (VEGAS) of 22 nearby clusters ($z < 0.05$), obtained an ICL luminosity fraction of $\approx 20\%(+15\%, -15\%)$. It should be noted that these estimated values are taken from empirical relations between the ICL fraction and cluster properties such as virial mass and the fraction of early-type galaxies (ETGs). However, the large scatter in these relations prevents a precise determination of the ICL fraction. 

     From the simulation perspective, \citet{Contini2014MNRAS} employed a semi-analytic model of galaxy formation and obtained an ICL mass fraction of $20$–$40\%$. In an updated study, \citet{Contini2023ApJ} used a state-of-the-art semi-analytic model and reported an ICL mass fraction of $ \approx 40\%(+10\%, -20\%)$. \citet{contreras2024characterising} analyzed simulated clusters from  The Three Hundred project and found an ICL mass fraction of $\approx 30\%(+10\%, -10\%)$. It is worth noting that \citet{Brough2024MNRAS} compared ICL fractions derived from different observational methods with the intrinsic values in simulated clusters,  and showed that current observational techniques tend to underestimate the intrinsic ICL fraction. For instance, when the ICL is measured using an aperture beyond $50$ kpc from the cluster centre, the intrinsic fraction can be underestimated by up to about $20\%$. 
    
     Combining the above observational and simulation results, in this work we adopt a fixed ICL luminosity fraction of $30\%(+20\%, -20\%)$ for clusters in our sample where direct observational measurements are not available. This assumed value and its uncertainty incorporate the observational data (about 20\%), as well as the simulated values (about 30–40\%) and the potential underestimation of the ICL claimed in the simulations.
     We use the average $M/L$ ratios of the galaxies in the canonical IMF or IGIMF cases to estimate the mass of the ICL in each galaxy cluster. The average $M/L$ ratio of the galaxies in a cluster is defined as the ratio between the total mass of all member galaxies and their total luminosity in a given photometric band. The uncertainty in the ICL mass is treated as a systematic error.

\section{Stellar mass boosting}
\label{SteMaBoost}
     In this section, the relation between $M/L$ ratios and luminosities of elliptical/spiral galaxies in the canonical IMF and IGIMF cases are shown for comparison. In addition to the B-band result depicted in Fig.~\ref{Bband}, Fig.~\ref{Kband},\ref{Vband},\ref{Rband} show the results in the K-, V- and r-band, respectively. The fit curves of the $M/L$ ratios were determined by cubic spline interpolation, and the uncertainties were derived from the error bound of the cubic spline interpolation \citep{hall1976optimal}.  
     
     These plots show that massive elliptical galaxies exhibit higher $M/L$ ratios across all bands under the IGIMF theory. For example, elliptical galaxies exhibit a rapidly increasing $M/L$ ratio if $\log_{10}(L_{\rm B}/L_{\rm \odot,B}) > 9.5$ in the B-band, ultimately reaching values approximately six times greater than those predicted by the canonical IMF.  In galaxy clusters, the cD/BCG and other high-luminosity galaxies are predominantly elliptical, which leads to a significant increase in their estimated masses under the IGIMF framework.
     As outlined in Sect. \ref{IGIMF}, this phenomenon arises because the IGIMF theory accommodates downsizing \citep{thomas2005epochs,yan2021downsizing}, namely that more massive (and luminous) elliptical galaxies undergo earlier and more rapid star formation (i.e., had higher star formation rates), resulting in an older stellar population and a higher proportion of stellar remnants compared to canonical IMF predictions. 
     The top-heavy gwIMF in the early phase is needed independently of the IGIMF theory in order to ensure rapid enrichment to super-solar metallicities.
     Consequently, for a given high-luminosity elliptical galaxy, the IGIMF theory predicts a higher mass than the canonical IMF, with this discrepancy becoming more pronounced as the luminosity of the galaxy increases. As a summary, these additional masses, compared to the canonical IMF, may either indicate an underestimation of the stellar mass itself due to the current bottom-heavy gwIMF at high metallicity, or be attributed to the presence of more stellar remnants, white dwarfs, and stellar-mass black holes.
     
     In contrast, spiral galaxies exhibit a minimal change in their estimated masses under the IGIMF theory. Due to their intrinsically lower luminosities and the smaller variations in their corresponding $M/L$ ratios, they are unlikely to affect the variation of total stellar mass of a galaxy cluster.
     
     Here we adopt the average $M/L$ ratio of cluster galaxies for the ICL. Since clusters hosting more massive elliptical galaxies tend to exhibit larger average $M/L$ ratios, the inferred ICL mass correspondingly becomes higher. 

     In summary, the masses of high-luminosity elliptical galaxies, particularly those in the central regions of clusters, as well as the total masses of clusters with a large fraction of massive ellipticals, are substantially increased when adopting the IGIMF theory.

\section{Results}
\label{Result}
     In this section we give the results derived from different datasets and compare them with relevant results documented in the literature.
\subsection{Comparison of clusters common in the WINGS and 2MASS catalogues}
\label{Sec_WINGS_2MASS_Common}
\begin{figure}
     \centering
     \includegraphics[width=0.45\textwidth]{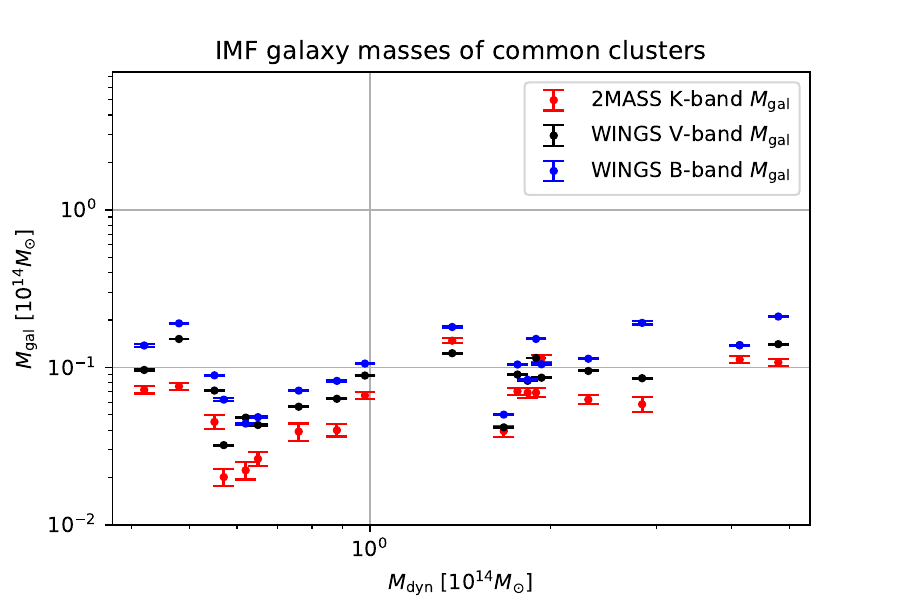}
     \includegraphics[width=0.45\textwidth]{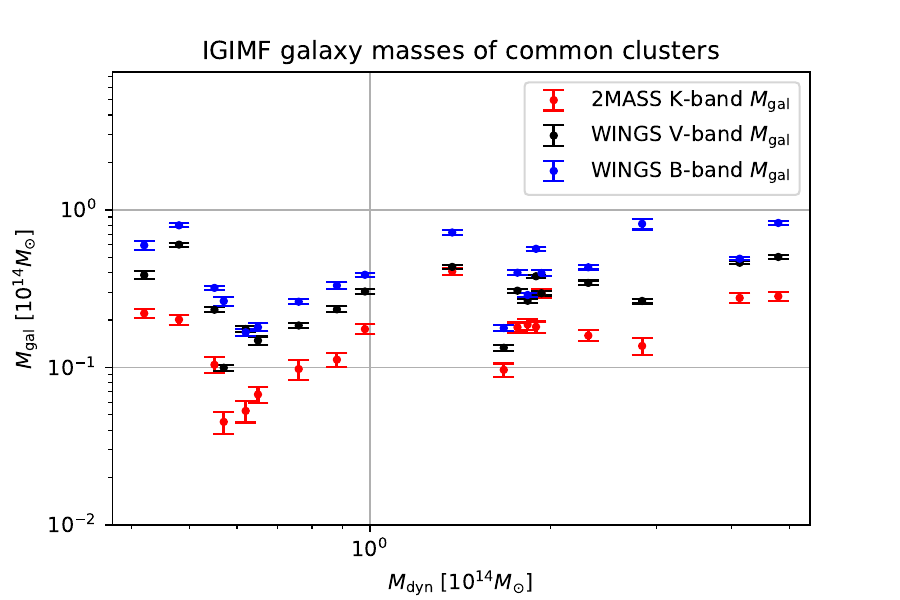}
     \caption{ Total galaxy masses, $M_{\rm gal}$, of 19 common clusters vs their MOND dynamical masses, $M_{\rm dyn}$, in the canonical IMF (upper panel) and IGIMF (lower panel) framework. Blue dots are in the WINGS B-band, black dots are WINGS V-band \citep{d2014surface,fasano2012morphology}, red dots are 2MASS K-band \citep{lin2004k,lin2004kk}. The auxiliary horizontal and vertical lines serve to emphasise the scale relationship relative to $10^{14}M_{\odot}$. Note that all MOND dynamical masses are from \citet{brownstein2006galaxy}.  }  
     \label{19_Common_IGIMF_IMF}
     \end{figure}

     Before applying the screening for spherical symmetry and hydrostatic equilibrium, we extracted 19 common galaxy clusters from both the WINGS (B- and V-bands) and 2MASS (K-band) surveys. To estimate the robustness of the $M/L$ ratios derived from the IGIMF theory, we focused on calculating the total galaxy mass for these clusters. The results are presented in Fig.~\ref{19_Common_IGIMF_IMF}.
     From the figure, it is evident that the B- and V-band results from WINGS in both IMF and IGIMF cases are quite similar. Additionally, the K-band results from 2MASS exhibit a similar overall distribution to the WINGS data, though with discrepancies in mass estimates. This difference is understandable, as the member galaxies used to fit the luminosity function in \citet{lin2004k} were restricted to those with absolute magnitudes $M_{\rm K} < -21$ (corresponds to the luminosity $ > 5 \times 10^{9} L_{\rm \odot,K}$), meaning only the more luminous galaxies were included. In contrast, the WINGS cluster samples include galaxies down to a fainter magnitude limit, significantly increasing the number of member galaxies considered within the same cluster compared to 2MASS. For instance, in the WINGS B-band data (\citet{d2014surface}), 751 and 530 member galaxies are identified in Abell 119 and Abell 133, respectively, whereas in 2MASS (\citet{lin2004k}) only $133 \pm 21$ and $150 \pm 40$ galaxies are recorded, respectively.      
     Another possible reason is the difference between observed and model-predicted colors. As shown in Fig.~\ref{color_compare}, the observed colors exhibit a large scatter, while our stellar population models reproduce only the central value of the color distribution.
     Furthermore, as noted in Sect. \ref{2MASSData}, some clusters may contain more than one BCG/cD galaxy, which further complicates mass estimation. Therefore, when using 2MASS data, we can only obtain a lower limit for the stellar mass of these 33 galaxy clusters discussed below.

     In conclusion, our analysis indicates the presence of a systematic uncertainty in the determination of cluster stellar masses. These masses are likely lower limits due to the incompleteness of the cluster catalogues.      
     And adopting the IGIMF results in stellar mass estimates that are typically a factor of 3–5 higher than those obtained with the canonical IMF.

\subsection{Results for the WINGS galaxy clusters}
\label{Sec_WINGS_result}
\begin{figure}
     \centering
     \includegraphics[width=0.45\textwidth]{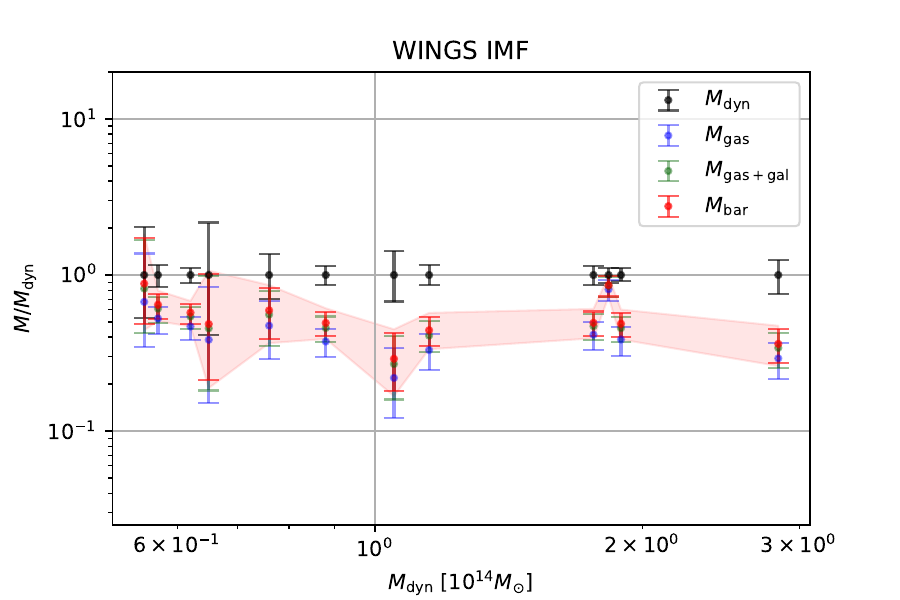}
     \caption{Mass relation of WINGS clusters in the canonical IMF case. The x-axis represents the MOND dynamical masses of the individual galaxy clusters \citep{brownstein2006galaxy}. The y-axis represents the individual mass components in units of the MOND dynamical mass. The black dots are the MOND dynamical masses, $M_{\rm dyn}$; the blue dots are the gas masses, $M_{\rm gas}$; the green dots are the gas + galaxy masses, $M_{\rm gas+gal}$; and the red dots are the gas + galaxy + ICL (total baryonic) mass, $M_{\rm bar}$ (error bars do not account for the ICL mass uncertainty). The auxiliary horizontal and vertical lines serve to emphasise the scale relationship relative to $M/M_{\rm dyn}$ and $10^{14}M_{\odot}$, respectively. The red shaded region represents the error range of total baryonic mass after including the systematic uncertainty from the ICL. All error bars are derived from the error propagation of $M/M_{\rm dyn}$.}
     \label{WINGS_IMF}
     \end{figure}

\begin{figure}
     \centering
     \includegraphics[width=0.45\textwidth]{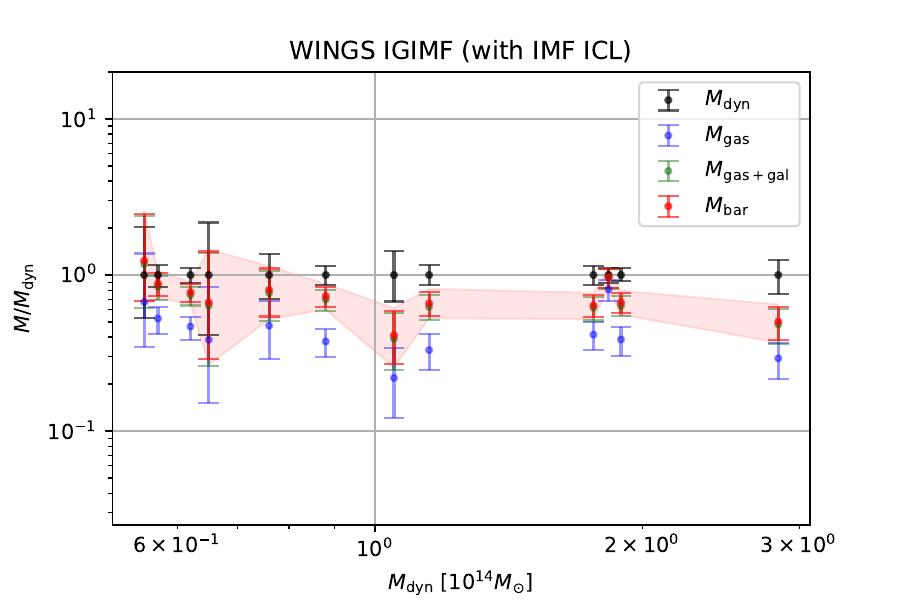}
     \caption{As in Fig.~\ref{WINGS_IMF}, but with galaxy masses calculated using the IGIMF and the ICL mass calculated using the canonical IMF.}
     \label{WINGS_IGIMF_ICLIMF}
     \end{figure}

\begin{figure}
     \centering
     \includegraphics[width=0.45\textwidth]{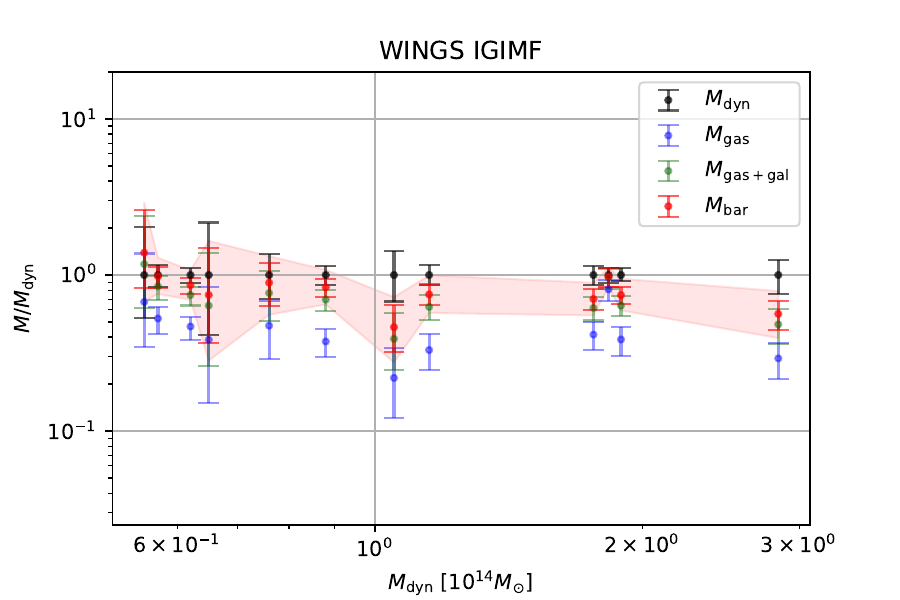}
     \caption{As in Fig.~\ref{WINGS_IMF}, but using the IGIMF to calculate all stellar masses.}
     \label{WINGS_IGIMF}
     \end{figure}

\begin{figure}
     \centering
     \includegraphics[width=0.45\textwidth]{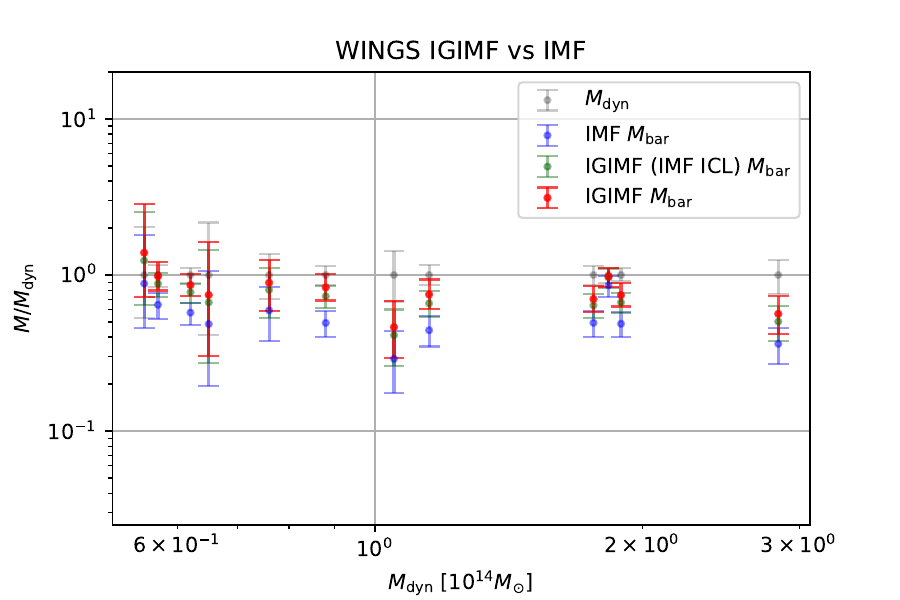}
     \caption{Relation between the total baryonic mass and the MOND dynamical mass of WINGS clusters for the canonical IMF, IGIMF with IMF ICL and IGIMF cases. The grey dots are the MOND dynamical masses, the blue dots are the total baryonic masses in the canonical IMF, the green dots are the total baryonic masses in the IGIMF with IMF ICL, the red dots are the total baryonic masses in the IGIMF framework. For simplicity, the systematic uncertainty from the ICL has been incorporated into the total baryonic mass error bars in all cases.} 
     \label{WINGS_IGIMF_vs_IMF}
     \end{figure}

     In this subsection, we present the mass results for the 12 galaxy clusters from the WINGS dataset. For the 12 galaxy clusters in the WINGS sample, we computed the mean values of the individual mass components based on their B- and V-band luminosities. The results are summarized in Table \ref{WINGS_table}. In the canonical IMF scenario, we present the MOND dynamical mass for each galaxy cluster, along with the corresponding gas mass, galaxy mass, ICL mass, and the total baryonic mass, which is the sum of the gas, galaxy, and ICL components. These results are illustrated in Fig.~\ref{WINGS_IMF}.
     The case of applying the IGIMF theory to calculate the galaxy mass while applying the canonical IMF to calculate the ICL mass is displayed in Fig.~\ref{WINGS_IGIMF_ICLIMF}, which can be seen as a middle case.
     The results for the IGIMF case are shown in Fig.~\ref{WINGS_IGIMF}.
     The comparison of the total baryonic mass in the three cases is plotted in Fig.~\ref{WINGS_IGIMF_vs_IMF}.
     
     As shown in Fig.~\ref{WINGS_IGIMF_vs_IMF}, the application of the IGIMF theory has significantly increased the stellar mass fraction in galaxy clusters. For most clusters, the total baryonic mass (with the systematic uncertainty) now falls within (or crossover of error bars) the 1$\sigma$ uncertainty range of the MOND dynamical mass, highlighting the potential of the IGIMF to address the missing baryonic mass problem in these systems.
     
\subsection{Results for the 2MASS galaxy clusters}
\begin{figure}
     \centering
     \includegraphics[width=0.45\textwidth]{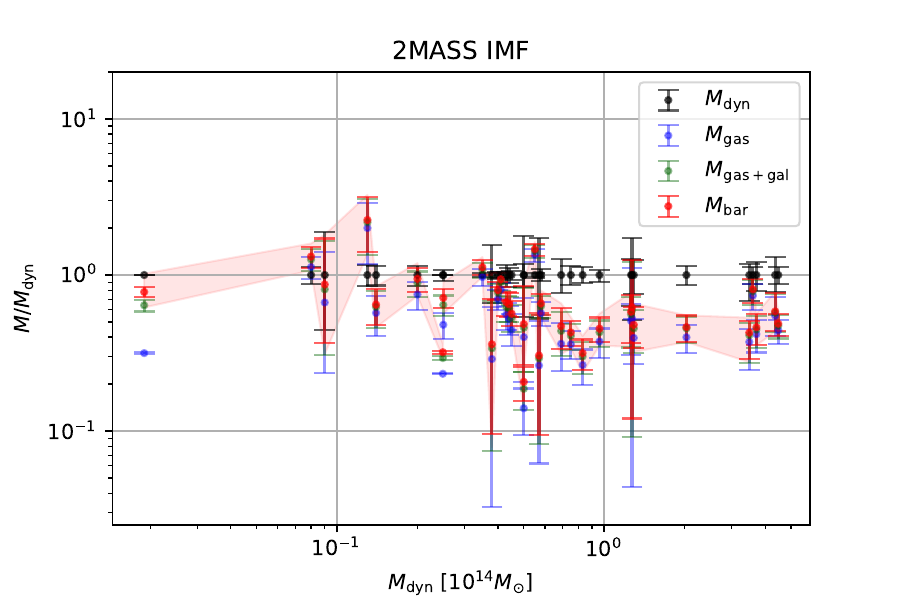}
     \caption{As Fig.~\ref{WINGS_IMF} but here for the 2MASS galaxy clusters.}
     \label{2MASS_IMF}
     \end{figure}

\begin{figure}
     \centering
     \includegraphics[width=0.45\textwidth]{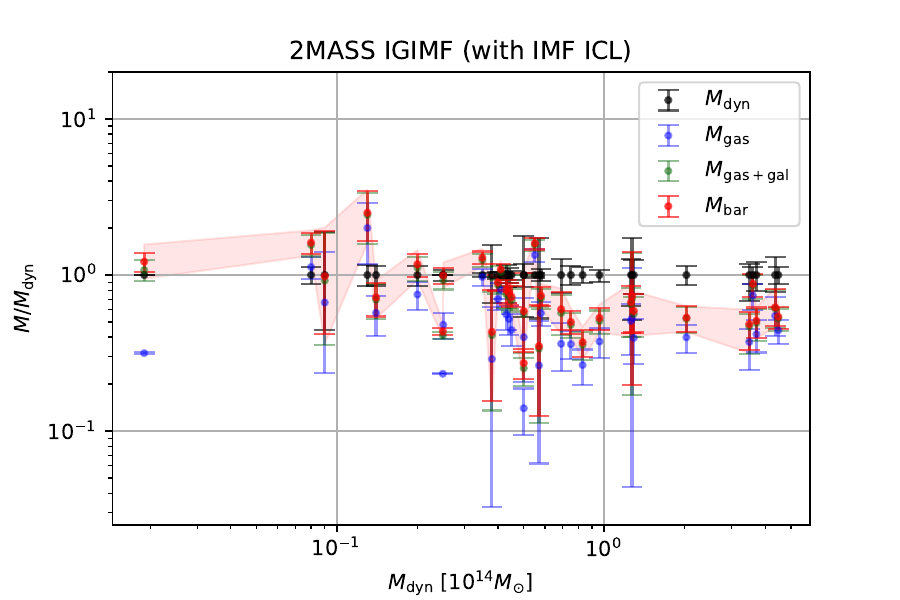}
     \caption{As Fig.~\ref{WINGS_IGIMF_ICLIMF} but here for the 2MASS galaxy clusters.}
     \label{2MASS_IGIMF_ICLIMF}
     \end{figure}

\begin{figure}
     \centering
     \includegraphics[width=0.45\textwidth]{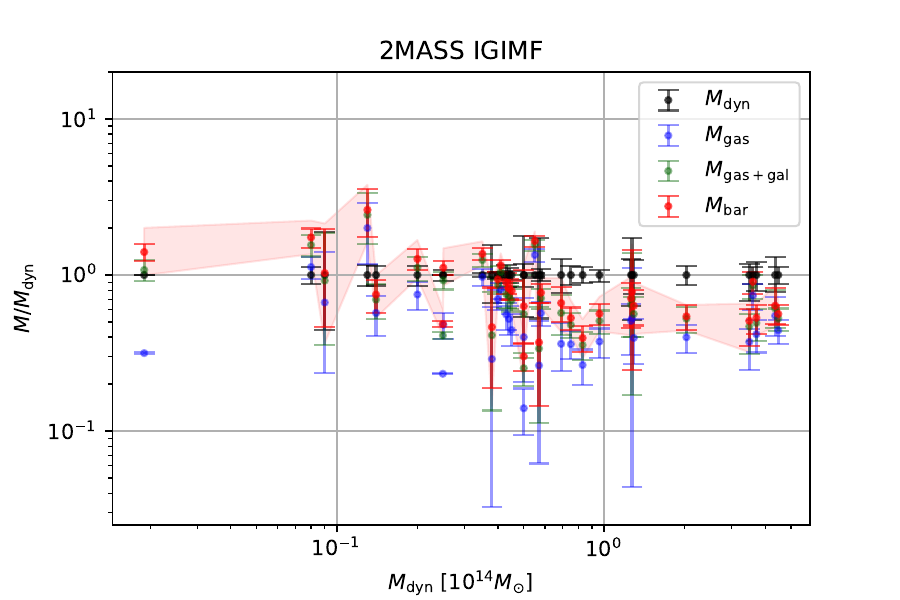}
     \caption{As Fig.~\ref{WINGS_IGIMF} but here for the 2MASS galaxy clusters.}
     \label{2MASS_IGIMF}
     \end{figure}

\begin{figure}
     \centering
     \includegraphics[width=0.45\textwidth]{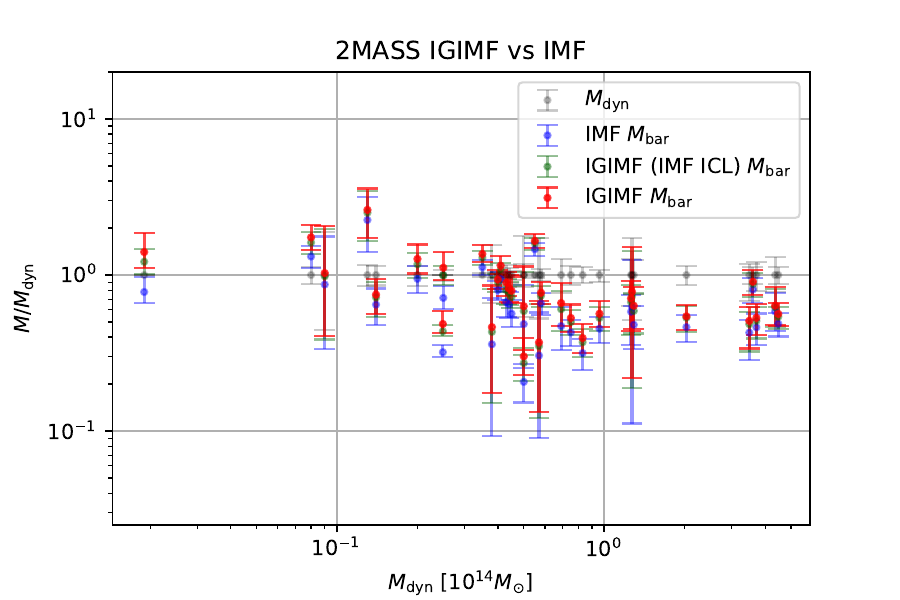}
     \caption{As Fig.~\ref{WINGS_IGIMF_vs_IMF} but here for the 2MASS galaxy clusters.}
     \label{2MASS_IGIMF_vs_IMF}
     \end{figure}

     In this subsection we show results for 33 galaxy clusters from the 2MASS dataset. The results are derived from K-band data reported in \citet{lin2004k} and are summarized in Table~\ref{2MASS_table}. As described in Sect. \ref{Sec_WINGS_result}, the results for the canonical IMF are presented in Fig.~\ref{2MASS_IMF}.
     The middle case of applying the IGIMF to calculate galaxy masses and the canonical IMF to calculate ICL masses is shown in Fig.~\ref{2MASS_IGIMF_ICLIMF}.
     The results for the IGIMF are presented in Fig.~\ref{2MASS_IGIMF}.
     The comparison of the total baryonic masses in these three scenarios is shown in Fig.~\ref{2MASS_IGIMF_vs_IMF}.
     
     Through comparison, it is evident that the galaxy clusters in the 2MASS dataset also exhibit significantly larger stellar mass components under the IGIMF theory. In more than half of the galaxy cluster cases studied here, the total baryonic masses (with the systematic uncertainty) derived using the IGIMF fall within (or crossover of error bars) or exceed the 1$\sigma$ uncertainty range of the MOND dynamical masses. Finally, as outlined in Sect. \ref{2MASSData} and Sect. \ref{Sec_WINGS_2MASS_Common}, we emphasize that the stellar masses (and therefore total baryonic mass) derived from the 2MASS data should be considered as lower limits, particularly for high-mass clusters, which typically host a large number of member galaxies.

\subsection{Comparison of all samples}
\begin{figure}
     \centering
     \includegraphics[width=0.43\textwidth]{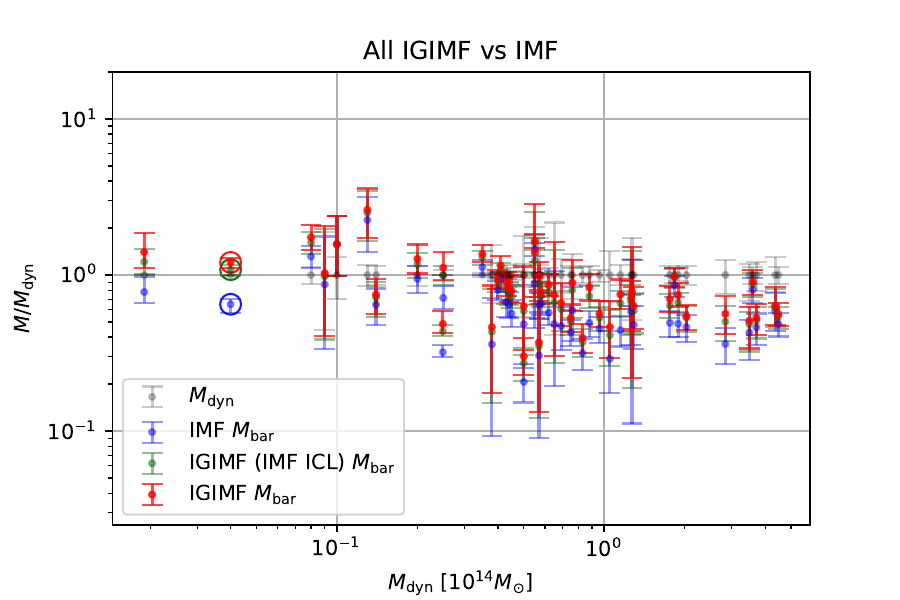}
     \includegraphics[width=0.43\textwidth]{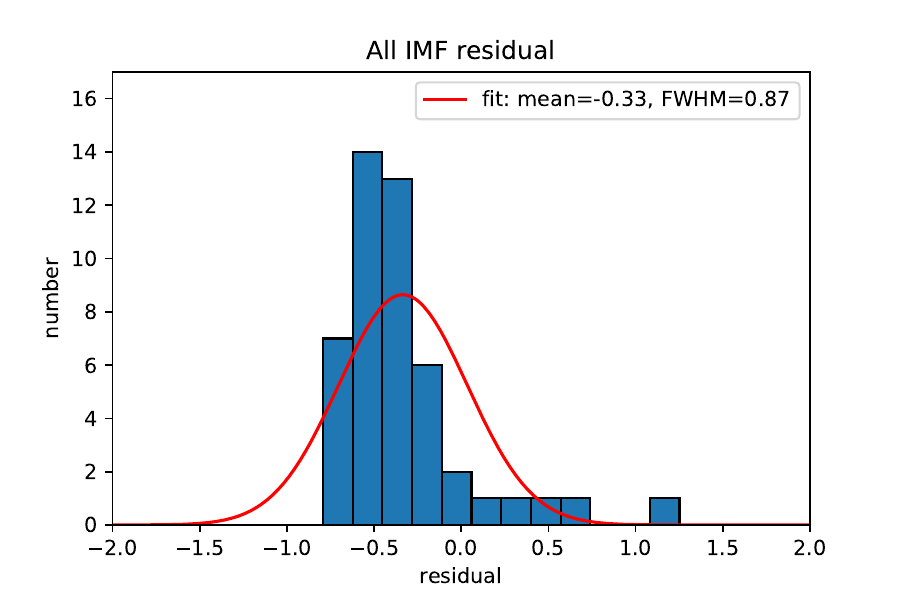}
     \includegraphics[width=0.43\textwidth]{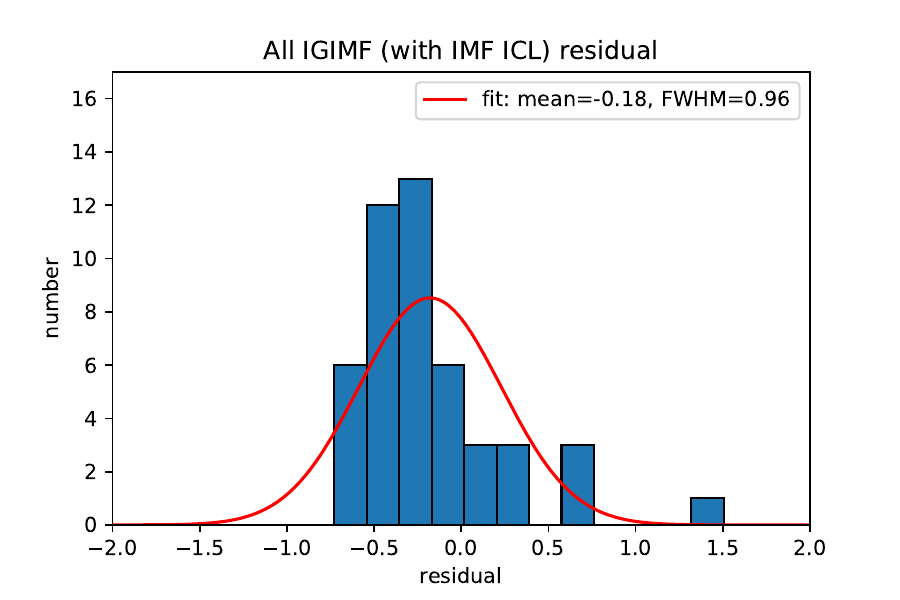}
     \includegraphics[width=0.43\textwidth]{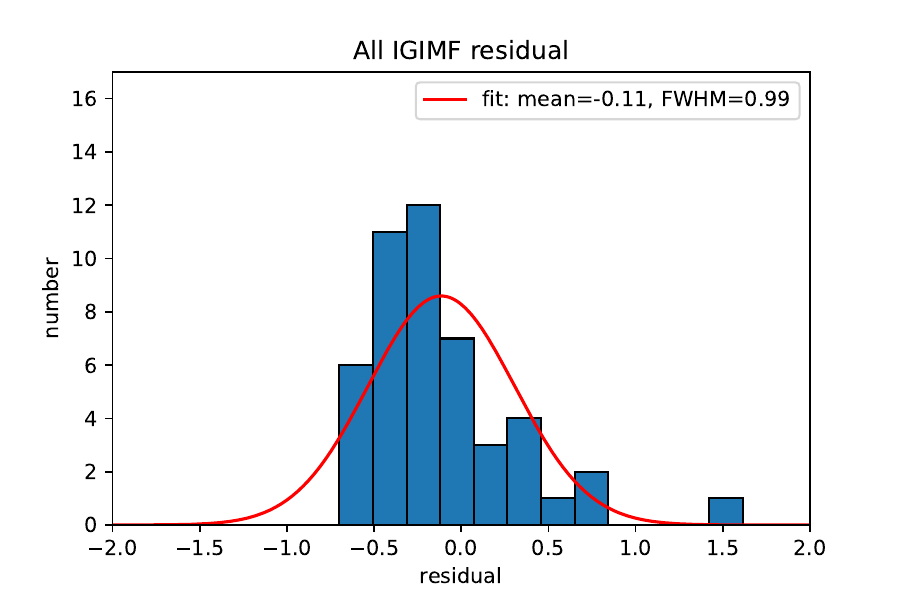}
     \caption{From top to bottom: Comparison of total baryonic masses for the three cases, with error bars including the systematic uncertainty from the ICL. Specially labeled circles indicate NGC 5044 (same color with corresponding mass component). Residuals are shown for the canonical IMF, the IGIMF with the ICL IMF, and the general IGIMF case. Blue histograms show the number distribution of residuals for the galaxy cluster sample; red curves represent Gaussian fits to the data; the mean and full width at half maximum (FWHM) are indicated.}
     \label{All_IGIMF_vs_IMF}
     \end{figure}

     Including the galaxy cluster NGC 5044 introduced in Sect. \ref{Sec_For_NGC}, we can sum up the sample of 46 clusters and give a complete comparison of the results in Fig.~\ref{All_IGIMF_vs_IMF}.  For NGC 5044, the mass estimates are derived solely from B-band data available in the \cite{mendel2008anatomy}.
     A clear outcome is that, by applying the IGIMF theory to more realistically assume the stellar population, the baryonic masses of nearly all sampled galaxy clusters are comparable to the MOND dynamical masses. The full dataset is presented in Tables \ref{WINGS_table}, \ref{2MASS_table}. Simple statistics for the overall sample show that the total baryonic mass contributes on average $67^{+4+2}_{-3-1}\%$ (the second error range accounts for the systematic uncertainty in the ICL mass) of the total MOND dynamical masses in the case of the invariant canonical IMF, $81^{+5+2}_{-4-1}\%$ in the case of the IGIMF (with IMF ICL), and $88^{+5+2}_{-4-1}\%$ in the case of the IGIMF. The gas masses, which are part of the baryonic masses, are not affected by the IMF/IGIMF and contribute a constant $52^{+4}_{-3}\%$ of the total MOND dynamical masses. 

     For NGC 5044, we find that the baryonic mass also reaches the dynamical mass in the IGIMF case. These results are indicated by the circles with corresponding colors in Fig.~\ref{All_IGIMF_vs_IMF}.
   
     We observe that in a few low-mass clusters, the total baryonic mass significantly exceeds the MOND dynamical mass. Aside from the dynamical uncertainties of hydrostatic equilibrium, this is likely due to the uncertainty in estimating the luminosities of cD/BCG and other bright elliptical galaxies. As noted by \citet{lin2004kk}, the luminosity fraction of cD/BCG galaxies decreases as the total cluster mass $M_{200}$ increases. That means, in low-mass clusters, cD/BCG galaxies can contribute a substantial fraction of the total stellar mass when using $M/L$ ratios predicted by the IGIMF theory. Therefore, the sensitivity of the $M/L$ ratio to the luminosity of these bright galaxies can amplify the overall uncertainty in the mass estimates. Uncertainties in the morphological classification of galaxies can further exacerbate this issue, as spiral galaxies misclassified as ellipticals may lead to significant overestimation of their stellar masses.

     We define the residual for each galaxy cluster as
     \begin{equation}\label{res}
     \begin{aligned} 
     Res_{i} = \frac{ M_{i \rm, bar} - M_{i \rm, dyn}  }{ M_{i \rm, dyn}  },
     \end{aligned}
     \end{equation}
     where the subscript $i$ denotes the $i$-th galaxy cluster. The corresponding plots of residuals for different cases are shown in Fig.~\ref{All_IGIMF_vs_IMF}. The residuals are fitted with a Gaussian distribution. We find that the residuals in the IGIMF case are more closely centered around zero compared to those in the canonical IMF case. This indicates that the baryonic masses predicted by the IGIMF theory provide a better fit to the MOND dynamical masses.

\subsection{Comparison of the radial mass profile for Abell 1795}
     \citet{Kelleher:2024oip} (hereafter \citetalias{Kelleher:2024oip}) investigated the spatial distribution of missing mass in galaxy clusters under the MOND framework, suggesting that the missing mass should be more concentrated than the ICM. Their study showed that within a radius of 200-300 kpc, the missing mass could be 1-5 times the known baryonic mass, while at distances of 2-3 Mpc, this ratio decreases to 0.4-1.1. We compare their findings with three galaxy clusters from our dataset that are in equilibrium: Abell 1795, Abell 2029, and Abell 2142. The comparison is summarized in Table \ref{3_table}.
     \begin{table*}
     \caption{Comparison of the three galaxy clusters from \citetalias{Kelleher:2024oip}, with all masses in units of $10^{14}M_{\odot}$.}             
     \label{3_table}      
     \centering                          
     \begin{ruledtabular}
     \begin{tabular}{c c c c c c c  c c}        
     Name & $M_{\rm b,200 kpc}$ & $M_{\rm miss,200 kpc}$  &  $\Delta M_{\rm BCG}$ & $r_{\rm out}$ [kpc]  &  $M_{\rm b,out}$  &  $M_{\rm miss,out}$  & $M_{\rm all,out}$  & $M_{\rm b,IG}$   \\    
     \hline                        
     A1795(no EFE) & 0.08 & 0.30  & $0.307^{0.033}_{0.033}$ & 2400 & $1.18^{0.295}_{0.295}$ & $0.896^{0.248}_{0.248}$ & $2.076^{0.385}_{0.385}$ & $1.871^{0.403}_{0.303}$    \\   
     A1795(EFE) & 0.065 & 0.30   &   &   &   & $1.298^{0.345}_{0.345}$ & $2.478^{0.454}_{0.454}$  &   \\
     A2029(no EFE)  & 0.18 & 0.30   & $0.086^{0.010}_{0.010}$ & 2700 & $2.42^{0.605}_{0.605}$ & $0.895^{0.296}_{0.296}$ & $3.315^{0.674}_{0.674}$  &  $2.624^{0.550}_{0.550}$\\      
     A2029(EFE)  & 0.15 & 0.38  &   &   &   & $1.476^{0.429}_{0.429}$ & $3.896^{0.742}_{0.742}$ & \\
     A2142(no EFE)  & 0.15 & 0.15  & $0.024^{0.004}_{0.004}$ & 2800 & $2.84^{0.710}_{0.710}$ & $0.738^{0.302}_{0.302}$ & $3.578^{0.772}_{0.772}$ & $2.983^{0.668}_{0.656}$  \\  
     A2142(EFE)  & 0.15 & 0.20  &   &   &   & $1.079^{0.384}_{0.384}$ & $3.919^{0.807}_{0.807}$ & \\
     \end{tabular}
     \end{ruledtabular} 
     \footnotesize{ From left to right: cluster name; visible baryonic mass within 200 kpc, $M_{\rm b,200kpc}$, from \citetalias{Kelleher:2024oip}; missing mass within 200 kpc, $M_{\rm miss,200kpc}$, from \citetalias{Kelleher:2024oip}; Mass boosting of BCG from the canonical IMF to the IGIMF cases (i.e., $\Delta M_{\rm BCG} = M_{\rm BCG,IG} - M_{\rm BCG,IMF}$, $M_{\rm BCG,IG}$ is the stellar mass of BCG in the IGIMF case); outermost radius $r_{\rm out}$, from \citetalias{Kelleher:2024oip}; visible baryonic mass within $r_{\rm out}$, $M_{\rm b,out}$, from \citetalias{Kelleher:2024oip}; missing mass within $r_{\rm out}$, $M_{\rm miss,out}$, calculated based on the missing mass fraction from \citetalias{Kelleher:2024oip}; $M_{\rm all,out} = M_{\rm b,out} + M_{\rm miss,out}$ is the required MOND mass within $r_{\rm out}$; the total baryonic mass within the virial radius assuming the IGIMF, $M_{\rm b,IG}$. Note that here we have included the ICL (together with its corresponding systematic uncertainty) and the galaxy masses of the corresponding clusters from Tables~\ref{WINGS_table} and \ref{2MASS_table}, with gas masses from \citetalias{Kelleher:2024oip}. EFE means the case with external field effect (specific to MOND), no EFE means isolated case.}      
     \end{table*}
     In this table, we show the mass boosting of the BCG from the canonical IMF to the IGIMF, $\Delta M_{\rm BCG}$, as well as the total baryonic mass within $r_{\rm out}$, $M_{\rm b,IG}$, obtained by applying the IGIMF. These should be compared to the missing mass within 200 kpc, $M_{\rm miss,200kpc}$, and the required total MOND mass, $M_{\rm all,out}$, from \citetalias{Kelleher:2024oip}, respectively.
     
     As seen in Table \ref{3_table}, $\Delta M_{\rm BCG}$ accounts for part of the missing mass within 200 kpc, whereas $M_{\rm b,IG}$ generally lies within the range of error bars for the required MOND mass, $M_{\rm all,out}$, within $r_{\rm out}$.

\begin{figure}
     \centering
     \includegraphics[width=0.45\textwidth]{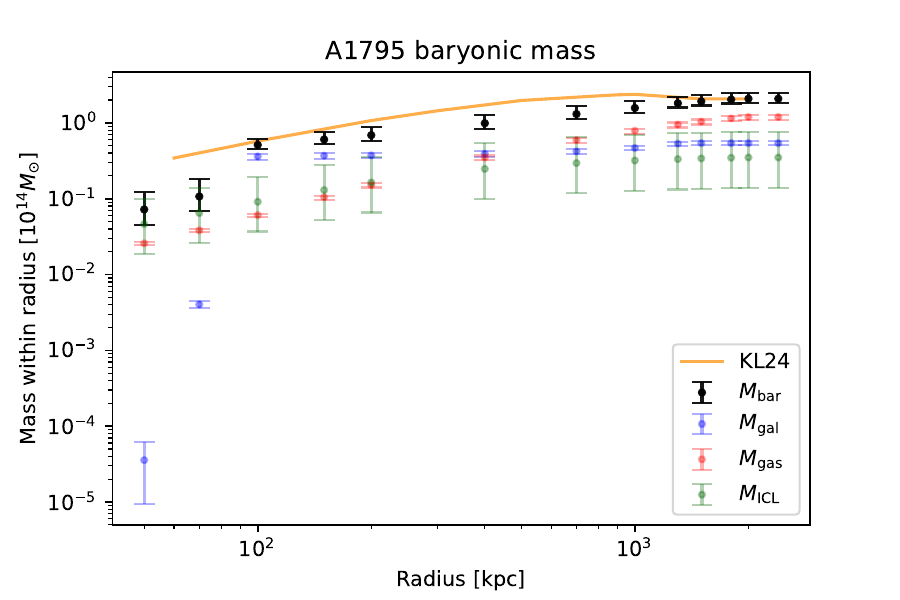}
     \caption{The total baryonic mass 2D projected distribution of A1795 is compared with the values from \citetalias{Kelleher:2024oip}. The \citetalias{Kelleher:2024oip} value (orange line) represents the sum of the observed baryonic and missing mass in the no-EFE case; the EFE case yields a similar result. The critical values are taken and converted from their Fig.~1. The IGIMF-based prediction (black dots, $M_{\rm bar}$) is the sum of the ICL (green dots, $M_{\rm ICL}$), galaxy (blue dots, $M_{\rm gal}$), and gas (red dots, $M_{\rm gas}$) mass components.} 
     \label{1795_vs_BV}
     \end{figure}

     For the three galaxy clusters listed in Table \ref{3_table}, we have only the galaxy catalog for Abell 1795 and have estimated the radial profile of its total baryonic mass.  
     In this calculation, we assumed that the ICL follows a King model \citep{sarazin1986x}, which is analogous to the $\beta$-model of the hot gas (adopted from \citet{brownstein2006galaxy}) but with $\beta \equiv 1$. The total gas mass, as well as the sum of the observed baryonic and missing mass, was taken from \citetalias{Kelleher:2024oip}.   
     The resulting profiles, converted into 2D projected distributions, are shown in Fig.~\ref{1795_vs_BV}.  As can be seen, the total baryonic mass predicted by the IGIMF is broadly consistent with the sum of the observed and missing baryonic components reported by \citetalias{Kelleher:2024oip}.

\section{Consistency of IGIMF-derived stellar masses with independent observational constraints}
\label{clarify}
      The IGIMF-based $M/L$ ratios adopted in this study, which include both stellar and remnant masses, predict significantly higher masses for luminous elliptical galaxies. This increase may lead to discrepancies when compared with independent mass estimates within the MOND framework, such as dynamical modeling and strong/weak gravitational lensing analyses. Several previous studies have shown that stellar masses derived using a canonical IMF are in good agreement with these independent measurements, whereas IGIMF-based stellar masses tend to exceed them. Even within the framework of Newtonian gravity, several studies have reported that elliptical galaxies may contain a relatively small dark matter component (see e.g., \citet{Cappellari2013MNRAS,Mamon2005MNRAS,Santucci2022ApJ}). In this section, we clarify and discuss this potential tension in MOND.

     First, massive elliptical galaxies have been found in numerous observations (\citet{2010NaturDokkum,2015ApJMart,2017ApJDokkum,2021A&AMart,2024MNRASBrok,2024ApJDokkum}) to possess a bottom-heavy gwIMF, which means that they contain a larger fraction of faint low-mass stars. This implies that using the $M/L$ ratios from a canonical IMF would underestimate their masses, making it necessary to re-examine cases where the baryonic masses of massive ellipticals inferred from a canonical IMF agree with MONDian dynamical or lensing masses. This does not imply that the baryonic masses reported in the above studies should simply exceed the dynamical or lensing masses. Dynamical and lensing masses themselves depend on modelling the mass distribution, which is in turn constrained by the observed baryonic distribution. Therefore, both the observational measurements and the modelling need to be jointly re-examined in light of the observational evidence for a bottom-heavy gwIMF.     
     In fact, there has already been a report showing that, within the MOND framework, high-luminosity ETGs have dynamical masses exceeding the baryonic masses estimated from a canonical IMF (\citet{2016MNRASDabringhausen}).

\begin{figure}
     \centering
     \includegraphics[width=0.45\textwidth]{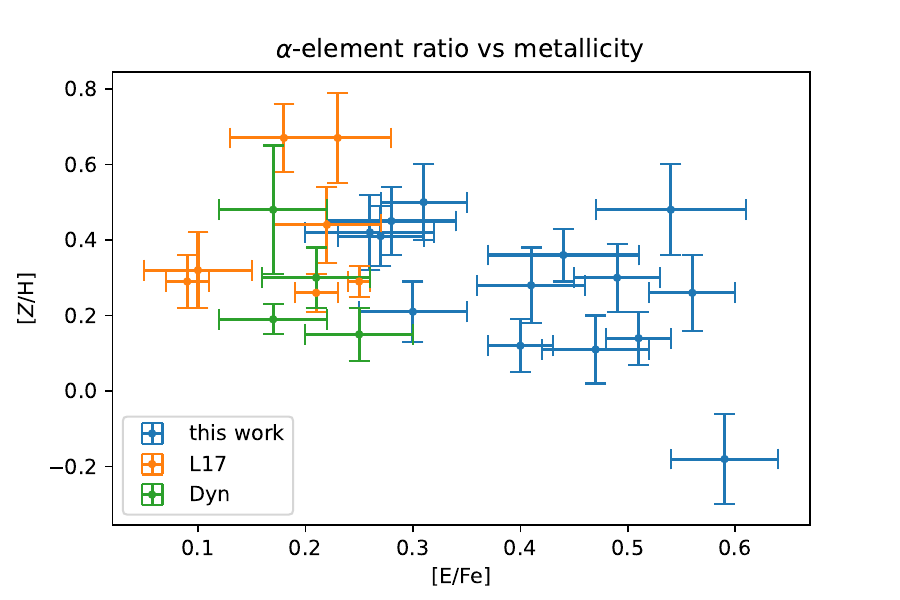}
     \caption{Comparison of metallicity and $\alpha$-element over Fe ratio for massive ETGs. Blue dots represent BCGs in several galaxy clusters from this work, with data taken from \citet{Loubser2009}. Orange dots correspond to X-ray ETGs from \citet{Lelli2017} (L17): NGC 4472 and NGC 4649 from \citet{Trager2000}, NGC 1521 from \citet{2007A&AAnnibali}, and NGC 720, NGC 1407, NGC 4125, and NGC 4261 from \citet{Denicol2005}. Green dots denote galaxies discussed in Sect.~\ref{Dyn_method} from several other studies (Dyn): NGC 821, NGC 2974, NGC 3379, and NGC 4494, with data taken from \citet{Denicol2005}. All measurements are taken within $r_{\rm e}/8$ or an approximate $r_{\rm e}/8$ aperture, where $r_{\rm e}$ is the effective radius of the galaxy.}
     \label{alpha_kband}
     \end{figure}

     Second, the IGIMF $M/L$ ratios for elliptical galaxies are sensitive to their stellar population age, formation history, and metallicity. In this work, we adopt the assumptions of monolithic collapse and downsizing, whereby massive elliptical galaxies form rapidly. These assumptions are supported by observations of elliptical galaxies (see \citet{2002MNRASChiosi,thomas2005epochs,2006ARA&ARenzini,yan2021downsizing,Jegatheesan2025A&A,2025NuPhBGjergo}). However, a subset of massive elliptical galaxies still exhibit relatively late star formation epochs and lower star formation rates.
     The star formation history of elliptical galaxies can be traced by their metallicity and the $\alpha$-element over Fe ratios. High $\alpha$-element ratios at high metallicity are strongly correlated with a short and intense period of star formation, and thus support the applicability of the IGIMF $M/L$ ratios used in this study. In Fig.~\ref{alpha_kband}, we present the metallicities and $\alpha$-element over Fe ratios (without distinguishing between $\alpha$-element ratio, [$\alpha$/Fe], and $\alpha$-enhancement ratio, [E/Fe], see e.g., \citet{Trager2000,Denicol2005}) for several sample galaxies whose canonical IMF-based stellar masses are consistent with MOND-based observational constraints. For comparison, we also show corresponding data for some BCGs in galaxy clusters studied in this work. 
     From Fig.~\ref{alpha_kband}, it can be seen that the BCGs in our cluster sample generally exhibit both high metallicity and high $\alpha$-element over Fe ratios. There are two galaxies from \citet{Lelli2017} with exceptionally high metallicities: NGC 720 and NGC 1407. However, according to the stellar population model from which we extracted the data \citep{Denicol2005}, both galaxies have stellar population ages of only $\approx$3 Gyr. Such young ages would lead to significantly lower IGIMF $M/L$ ratios (see e.g., \citet{Zonoozi2025}). The derived elemental abundances and ages are model-dependent, and different models may yield different results. In \cite{thomas2005epochs,Humphrey2006}, the estimated ages of these two galaxies are higher, but their reported metallicities are also substantially smaller.

\begin{figure}
     \centering
     \includegraphics[width=0.45\textwidth]{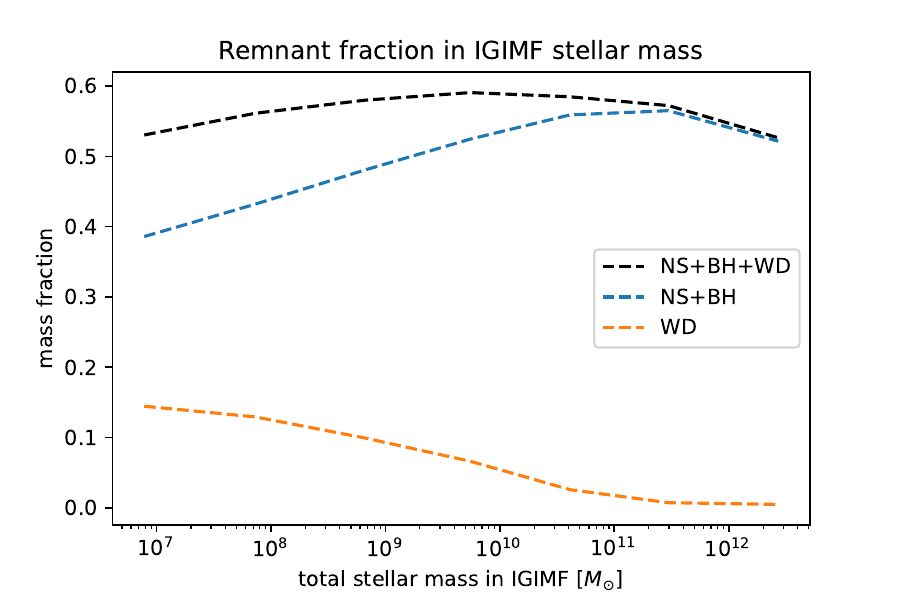}
     \includegraphics[width=0.45\textwidth]{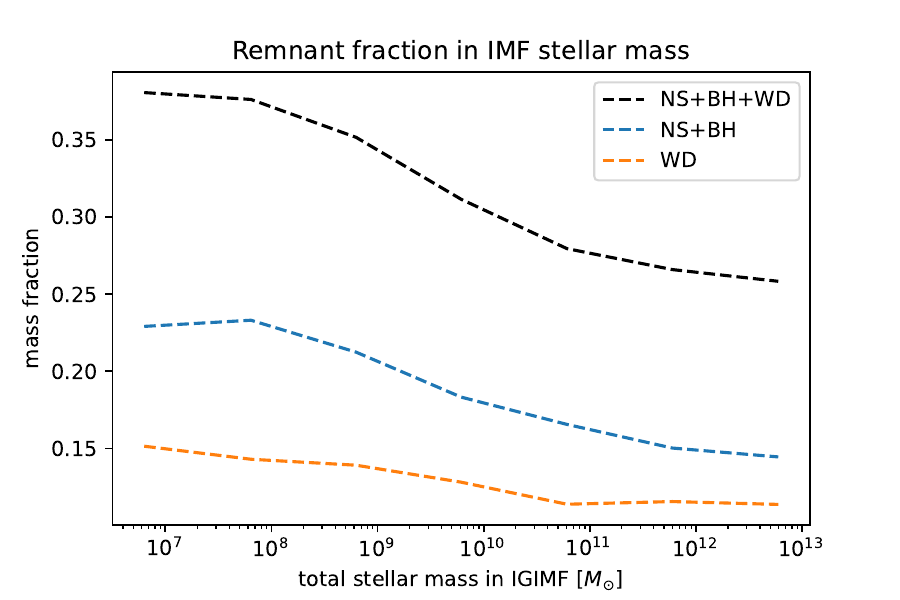}
     \caption{Total stellar mass (including remnants) of ETGs as a function of the mass fraction contributed from different stellar remnants, shown for the IGIMF (top panel) and canonical IMF (bottom panel) cases, assuming a stellar population age of 12 Gyr. The mass fractions of neutron stars (NSs) and stellar-mass black holes (BHs) are indicated by the blue dashed line, while that of white dwarfs (WDs) is shown by the orange dashed line. The total remnant mass fraction (NS + BH + WD) is indicated by the black dashed line. Note that for massive ETGs, the IGIMF stellar masses are about four to six times higher than those inferred from a canonical IMF.}
     \label{remnants}
     \end{figure}

     Finally, as discussed above, previous studies suggest that massive elliptical galaxies underwent early and rapid star formation. After a long period of stellar evolution, these galaxies are expected to contain a significant population of stellar remnants (white dwarfs, neutron stars, stellar-mass black holes), along with an increased fraction of low-mass stars (that is, these galaxies had a top-heavy gwIMF at the time of their initial formation, which was needed to rapidly evolve the star-forming gas to a high metallicity, and now exhibit a bottom-heavy gwIMF as a consequence of their high metallicity; see e.g., \citet{2024arXiv241007311K}; also \citet{jevrabkova2018impact,yan2021downsizing,Haslbauer2024,Zonoozi2025}).     
     The mass of these remnants and low-mass stars is included and the resulting high $M/L$ ratios reflect this feature.   
     However, the spatial distribution of these remnants remains an open question that requires further investigation. 
     For example, recent studies have shown that neutron stars can receive substantial kick velocities \citep{2025arXivDisberg}, with a peak around $200$~$\mathrm{km/s}$ relative to the local standard of rest (LSR) of pulsars. Considering that the progenitor stars of these neutron stars originated from the Milky Way disk and therefore had relatively low velocities with respect to the LSR, the measured velocities can be approximated as relative to the progenitor stars. Given that the typical velocity dispersion of massive elliptical galaxies is about $200$--$300$~$\mathrm{km/s}$ (see e.g., \citet{2014MNRASLyskova}), the velocities of neutron stars relative to the LSR in ETGs should be approximately the progenitor velocity and the kick velocity added in quadrature (and possibly even higher if a galaxy merger history is present). This implies that the spatial distribution of remnants is more extended than that of the stars, so that a significant fraction of neutron stars and stellar-mass black holes may reside beyond the galaxy's effective radius.      
     Consequently, the $M/L$ ratios within the effective radius could remain closer to that predicted by a canonical IMF. Many independent mass estimates are derived specifically within this inner region, potentially explaining the consistency observed with canonical IMF-based models. In Fig.~\ref{remnants}, we show the relation between ETG stellar masses and the fractional contributions of stellar remnants in the IGIMF and canonical IMF cases. Note that in the IGIMF framework, the stellar masses of massive ETGs are four to six times higher than those inferred from a canonical IMF. The figure clearly shows that in the IGIMF case, stellar remnants in massive ETGs are dominated by neutron stars and stellar-mass black holes, accounting for approximately 50\% of the total stellar mass of the galaxy, whereas in the case of a canonical IMF they contribute only about 15-20\%. The decrease of the fraction of remnant mass with increasing ETG mass for the canonical IMF comes from less-massive ETGs having a lower metallicity which leads to more massive remnants.  
     
     It should be noted that when the mass fraction of black holes is significantly larger than that of neutron stars, the explanatory effectiveness of kick velocity arguments may be reduced, since from a theoretical perspective black holes are expected to receive smaller kick velocities than neutron stars in order to satisfy momentum conservation. Nevertheless, a variety of observational constraints influence our knowledge of the kick velocity distribution of black holes. For example, present-day-type globular clusters (GCs) may have hosted a more top-heavy IMF \citep{Marks2012}, and studies by \citet{Peuten2016} and \citet{Baumgardt2017} suggest that approximately 50\% of the total number of black holes can be retained in GCs (see also \citealt{Pavlik2018}). This implies that the total mass of the black-hole-hosting GCs could be 10--100 times larger than thought at present, which would require correspondingly larger kick velocities for black holes to escape from the deeper potential wells of the young GCs. In addition, the relative fractions of implosion and explosion outcomes in progenitors remain uncertain \citep{Wirth2024}, leading to uncertainties in the fallback mass fraction and thus in the resulting kick velocity distribution. The metallicity of progenitors may also affect black hole kick velocities; massive ETGs experience rapid metal enrichment, resulting in a wide metallicity range for progenitors. Observational constraints on black hole kick velocities are currently mainly derived from binaries, as isolated stellar-mass black holes are difficult to detect and high kick velocities may disrupt their host binaries, if present\citep{Nagarajan2025}. Moreover, black hole–star binaries may form in star clusters through three-body encounters, such that these systems thus only provide indirect constraints on the kick velocity distribution function of black holes.
     
     Another potential issue concerns projected mass effects. Considering the strong-lensing 2D projected mass enclosed within the Einstein radius, $R_{\rm E}$, of a galaxy, 
     \begin{equation}
     M_{\rm 2D}(<R_{\rm E}) = \pi R_{\rm E}^{2}\,\Sigma_{\rm critical},
     \end{equation}
     where $\Sigma_{\rm critical}$ is the critical surface density, and $R_{\rm E}$ is often close to the galaxy effective radius.
     This projected mass includes contributions from matter that lies within the projected radius $R_{\rm E}$ but is located at larger distances along the line-of-sight.
     If the inferred lensing mass remains consistent with the predictions of a canonical IMF even after accounting for such projection effects, the explanatory effectiveness of the kick velocity argument may be weakened.
     However, the contribution of stellar remnants to the projected mass depends on the fraction of their mass enclosed within the projected radius, which is reduced if kick velocities distribute the remnants more broadly.
     As a simple example, consider a model in which stellar remnants are uniformly distributed on a thin spherical shell of radius $R_{\rm shell} = R_{\rm E}$ centered on the galaxy. In this case, the projected strong-lensing mass within $R_{\rm E}$ includes the full mass of the shell. If kick velocities expand the shell to $1.25R_{\rm E}$, only 40\% of the remnant mass remains projected within $R_{\rm E}$. For a expansion to $1.5R_{\rm E}$, this fraction further decreases to 25.5\%.
     The total baryonic mass enclosed within the projected radius in the IGIMF$+$kick-velocity scenario may become comparable to that obtained under a canonical IMF without kick velocities (which is the assumption commonly adopted in most previous studies), and thus may provide a viable explanation for the MOND strong-lensing mass.     
     A more thorough and quantitative investigation of this issue will require combining lensing constraints with spatially resolved mass distribution models.
     
     The \texttt{SPS-VarIMF} code is currently under development to estimate the individual mass fractions and mass functions of black holes and neutron stars. 
     On the other hand, stellar remnants may contribute a substantial fraction of the total mass in ETGs, making it unavoidable to perform self-consistent galaxy-cluster-scale assembly simulations with sufficient resolution in order to study the interplay between stellar-remnant dynamics and the evolving global potential of galaxy clusters.
     By combining existing observational constraints with simulations of spatial dynamics (note that no MOND-based hydrodynamical codes currently implement a varying stellar IMF yet), these efforts are expected to enable future studies to model the spatial distributions of stellar remnants within the IGIMF framework.

     In summary, this tension may be alleviated or clarified through several possible avenues. First, observational evidence for a bottom-heavy gwIMF in massive elliptical galaxies conflicts with previous findings that baryonic masses based on a canonical IMF agree with MOND dynamical or lensing masses, indicating the need for a reassessment of this issue. Second, the IGIMF $M/L$ ratios depend on multiple properties of the stellar populations. The $M/L$ assumptions adopted in this work are based on observations of elliptical galaxies and are suitable for estimating the masses of highest-luminosity ellipticals. For galaxies where canonical IMF $M/L$ ratios match observations, more realistic IGIMF $M/L$ ratios informed by elemental abundances could help relieve the tension. Finally, the IGIMF predicts a large amount of stellar remnants, but their detailed spatial distribution remains to be investigated. Recent studies support the possibility of a more extended distribution. Constructing spatially resolved IGIMF $M/L$ ratios based on the stellar population characteristics of individual galaxies, combined with dynamical and lensing simulations, would improve the predictive power of the IGIMF theory for different galaxy types. However, this is beyond the scope of the present work and will be addressed in future studies.

     In the following, we discuss each case according to the observational method used.

\subsection{The strong and weak gravitational lensing mass}

     For a long time, there has been tension between the MOND Theory and (strong/weak) gravitational lensing observations (e.g., \citet{gavazzi2002constraints,takahashi2007weak,natarajan2008mond,ferreras2008necessity,tian2009relation}). A key issue is that MOND lensing masses are typically larger than the total mass of visible baryons within galaxies and the core area of galaxy clusters. Without modifying the fundamental MOND formalism, this discrepancy must be explained by invoking additional mass sources. Meanwhile, several studies have reported that the baryonic masses of galaxies assuming a canonical IMF are consistent with the masses inferred from strong or weak lensing (e.g., \citet{2017MNRASTian}). In this context, we emphasize that it is crucial to consider the radial range probed by the lensing observations. 

     In galaxy-scale strong lensing systems, the Einstein radius typically lies close to the galaxy’s effective radius. As discussed above, due to the large kick velocities of neutron stars, a significant fraction of these high-$M/L$ ratio remnants may lie beyond the effective radius. As a result, the $M/L$ ratio within the effective radius may remain close to that expected from a canonical IMF, which could explain the agreement between strong lensing masses and canonical IMF-based baryonic masses in such cases (see e.g., \citet{2017MNRASTian}).

     In contrast, in studies such as \cite{gavazzi2002constraints,ferreras2008necessity}, where the Einstein radius extends well beyond the effective (or half-light) radius, the authors report a discrepancy between the mass inferred from lensing and that estimated assuming a canonical IMF.

     Moreover, the lensing systems are mostly located at higher redshifts compared to the sample analyzed in this study. It is worth noting that at redshifts $z > 0.1$, elliptical galaxies are typically younger and more luminous, resulting in slightly lower $M/L$ ratios than those discussed in Sect. \ref{SteMaBoost}. The V- and r-band $M/L$ ratios for elliptical galaxies at $z = 0.3$ are shown in Fig.~\ref{Vband_0.3} and Fig.~\ref{rband_0.3}, and can be used for comparison with those in Fig.~\ref{Vband} and Fig.~\ref{Rband}, respectively. At higher redshifts, for example in some galaxy cluster samples from the Cluster Lensing And Supernova survey with Hubble (CLASH) project \citep{2012ApJPostman} with redshifts reaching $z \approx 0.6-0.8$, the IGIMF framework still predicts that massive ETGs have $M/L$ ratios that are a factor of 2--3 higher than those inferred under a canonical IMF in the corresponding bands. This is because a substantial fraction of the stellar remnant mass has already been accumulated during the early stages of galaxy formation, as discussed in Section~\ref{IGIMF}.

     For instance, \citet{gavazzi2002constraints} provide constraints on the galaxy cluster MS2137-23 based on strong and weak lensing effects. At the Einstein radius of the core area of MS2137-23, $r_{\rm E} = 58$ kpc, the mass of the X-ray gas is approximately $4\times 10^{12} M_{\odot}$, while the lensing constraint on the MOND mass is around $2\times 10^{13} M_{\odot}$. Considering the BCG of this cluster, with a V-band absolute magnitude of $M_{\rm V} = -24.38 \pm 0.09$, redshift $z=0.313\pm0.001$ and an effective radius of $r_{\rm e,V} = 24.8 \pm 0.34$ kpc \citep{sand2002dark}, we estimate the BCG mass in the IGIMF theory to be $1.36\times 10^{13} M_{\odot}$. (V-band stellar mass-to-light ratio $M/L_{\rm \odot,V} \approx 28$ for $L_{\rm V} \approx 4.84 \times 10^{11} L_{\rm \odot,V}$, see Fig.~\ref{Vband_0.3}). This reduces the gap between the MOND lensing mass and the baryonic mass to about 10\%. Since the author did not provide an uncertainty range for the MOND lensing mass estimate, we can only make a qualitative statement that, under the IGIMF theory, there is no noticeable discrepancy between the baryonic mass and the MOND lensing mass within the Einstein radius.

     In contrast to strong lensing, weak lensing probes a much larger region of the galaxy, and thus does not exclude a significant fraction of stellar remnants. One example is provided by \citet{tian2009relation}, who established a relation between the $M/L$ ratio required by MOND and the r-band luminosity of galaxies, based on weak lensing mass constraints and observed galaxy luminosities. Their results show that more luminous galaxies require higher $M/L$ ratios, with values reaching $M/L = 10$–$20$ for luminosities around $10^{10}L_{\odot,r}$. Since their study primarily focused on ETGs (mainly elliptical galaxies), the IGIMF theory naturally explains the elevated $M/L$ values in such high-luminosity systems (see Fig.~\ref{Rband} and Fig.~\ref{rband_0.3} for the r-band $M/L$ ratios).

\subsection{The radial acceleration relation}     

     The radial acceleration relation (RAR; see, e.g., \citet{McGaugh2016PhRvL,Lelli2017}) describes the correlation between the observed radial acceleration in a galaxy and the acceleration estimated from its baryonic mass, which is inferred from the luminosity assuming a canonical IMF and the gas mass. This relation constitutes a central prediction of MOND.     
     Given that the $M/L$ ratios derived from the IGIMF theory for high-luminosity ETGs are significantly larger than those inferred from the canonical IMF, it is essential to discuss whether this discrepancy leads to a deviation of these galaxies from the RAR.

     In \citet{2023MNRASDabringhausen}, the authors have shown that ETGs in the IGIMF framework match the MONDian RAR at the half-light radius. To extend the RAR analysis to larger radii, weak lensing has been employed. In \citet{2021A&ABrouwer}, the authors indeed found that the observed acceleration of ETGs from weak lensing, when combined with the baryonic acceleration based on a canonical IMF, does not follow the RAR, requiring roughly a factor of two baryonic mass to bring them onto the relation. Subsequently, \citet{2024JCAPMistele} improved upon the method of \cite{2021A&ABrouwer} and reported that the weak-lensing-based acceleration of ETGs is consistent with the RAR. This appears to be in tension with the higher baryonic masses suggested by the IGIMF in this work. However, it should be noted that the stellar masses of the sample in \cite{2024JCAPMistele}, based on a canonical IMF, were capped at $10^{11} M_{\odot}$, a range where the IGIMF $M/L$ ratios are only about a factor of two higher than those of the canonical IMF. Moreover, the galaxies in \cite{2024JCAPMistele} are at redshifts $z = 0.1$-$0.5$, where the IGIMF $M/L$ ratio decreases (see Fig.~\ref{Vband_0.3},\ref{rband_0.3}). The method used in \cite{2024JCAPMistele} to estimate the baryonic mass of ETGs was also modified compared to \cite{2021A&ABrouwer}: bright ETGs were assigned stellar masses about 40\% higher than in \cite{2021A&ABrouwer}, and, in addition, X-ray gas masses were added based on scaling relations. For ETGs at the upper mass limit of $10^{11} M_{\odot}$, the extra gas mass can reach $\approx 60\%$ of the stellar mass. Therefore, the massive elliptical galaxies in \cite{2024JCAPMistele} were essentially matched to the RAR using baryonic masses already higher than those from a canonical IMF.

     On the other hand, the IGIMF $M/L$ ratio reflects the star formation history implied by the elemental abundances of galaxies. In the work of \cite{Lelli2017}, the X-ray ETGs studied are located in small galaxy groups or relatively isolated environments, whereas in this study we focus on galaxy clusters with high-density environments. ETGs in low-density environments form significantly later than those in high-density environments \citep{thomas2005epochs}, which can affect the degree of stellar remnant accumulation. Our sample BCGs generally exhibit higher metallicities and $\alpha$-element ratios (Fig.~\ref{alpha_kband}). This supports the scenario in which the initial stellar populations had a top-heavy gwIMF, resulting in a higher present-day mass fraction of stellar remnants. 
     In fact, BCGs have been observed to deviate from the standard RAR in both dynamical and gravitational lensing measurements \citep{2020ApJTian,2024A&ATian}, with the fraction of missing mass increasing as the acceleration decreases. 
     An independent study \citep{2024MNRASBrok} also found that BCGs and their massive satellite galaxies (mostly high-luminosity ETGs) exhibit similar elemental abundances, supporting the idea that their early star formation proceeded with a top-heavy gwIMF, while their present-day stellar populations follow a bottom-heavy gwIMF. This implies that, in terms of the RAR, massive satellite galaxies are likely to behave similarly to BCGs, namely showing deviations from the standard RAR.
     This suggests that galaxies in cluster environments may have different properties from those in isolation. 
     Such differences should be reflected in the IGIMF model, a topic that lies beyond the scope of the present work and will require further investigation in the future.
 
\subsection{The dynamical method}
\label{Dyn_method}

     Several studies have also reported that massive elliptical galaxies show good agreement between their dynamical masses and the baryonic masses inferred with a canonical IMF (see e.g., \citet{2003ApJMilgrom,2007A&ATiret,2012PhRvLMilgrom,2017MNRASTian}). Except for \cite{2017MNRASTian}, whose measurements are confined to within the effective radius, the other studies extend beyond the effective radius. For comparison, we show in Fig.~\ref{alpha_kband} the elemental abundances of these systems, where NGC 720 and NGC 1521 are also included in the sample of \cite{Lelli2017}. As seen in the figure, their $\alpha$-element ratios are generally lower than those of the BCGs in our sample. This implies that their star formation occurred later and proceeded more slowly, leading to $M/L$ ratios that are closer to those expected from a canonical IMF. The detailed correlation between $M/L$ ratio values, luminosities, and elemental abundances awaits further research, which is now possible within the IGIMF theory by using the codes published by \cite{Haslbauer2024} and \cite{Zonoozi2025}.

\subsection{The gravitational wave observatories}     

     Reaching the observed super-solar metallicities on Gyr timescales \citep{yan2021downsizing} requires massive ETGs to host a larger number of stellar remnants, which may impact the rates of neutron star–black hole and black hole–black hole mergers detectable by gravitational-wave observatories. An important question is therefore whether the currently observed merger rates already exclude such an increased remnant population. While the number of detected merger events is increasing, current population-synthesis models still need to be calibrated against these observations. The predicted merger rates remain uncertain, in part because the relative contributions of remnant binaries formed through three-body encounters in star clusters versus neutron star--black hole and black hole--black hole binaries originating directly from binary stellar evolution are not well constrained. This balance depends not only on the birth properties of binary star systems, but also on the densities of newly formed star clusters, within which stellar-dynamical encounters can rapidly reshape an initial binary population \citep{Kroupa2025CoSka}. In addition, the formation rates of star clusters in different galaxies, the properties of their constituent stellar populations, and the assumed shape of the stellar IMF in population-synthesis models (e.g. \citealt{Mapelli2019,Banerjee2022}) are all uncertain and may require revision. Moreover, when kick velocities are taken into account, high kick velocities can unbind binaries, thereby suppressing the formation of merging compact-object binaries. Consequently, given the current observational data and theoretical uncertainties, it remains difficult to rule out the presence of a large population of stellar remnants.

\section{Conclusion}

     In this work, we have recalculated the stellar masses of galaxy clusters using the $M/L$ ratios derived from a realistic calculation of stellar population using the IGIMF theory, providing updated estimates for their total baryonic mass.    
     Due to the top-heaviness of the galaxy-wide IMF as calculated using the IGIMF theory and as needed in order to synthesize the super-solar metallicities of the massive central elliptical galaxies \citep{yan2021downsizing}, the $M/L$ ratios of elliptical galaxies are significantly larger than previously thought in the context of an assumed invariant stellar IMF. Because they typically constitute a significant proportion in the population of galaxies in galaxy clusters, both the total stellar mass and the BCG mass (as well as the ICL mass of high-mass galaxy clusters) in the core regions become significantly higher.        
     The mean ratio of the total baryonic mass compared to the MOND dynamical mass increases from $67^{+4+2}_{-3-1}\%$ in the canonical IMF case to $88^{+5+2}_{-4-1}\%$ in the IGIMF case, meanwhile the ICM provides a constant mean ratio of $52^{+4}_{-3}\%$.
     Our findings are supported by the analysis of a sample of 46 selected galaxy clusters \citep{fasano2012morphology,d2014surface,lin2004k,lin2004kk, angus2008x, brownstein2006galaxy,mendel2008anatomy}. 
     
     Our estimate of $88^{+5+2}_{-4-1}\%$ is likely an underestimation because it is based on the known census of galaxies in the clusters, which is likely to be incomplete. Overall, our estimation method is conservative. 
     Furthermore, since all the datasets used have lower limits on magnitude or brightness, a large number of faint spiral galaxies with high gas fractions may not have been recorded, also potentially explaining part of the missing baryons (see e.g., \citet{binggeli1988luminosity,mcgaugh1997gas}).
     
     We compared our results with a recent study on MOND missing masses (\citetalias{Kelleher:2024oip}). In particular, we constructed the baryonic mass profile of Abell 1795 under the IGIMF assumption and compared it with the profile of the observed mass plus the missing mass as reported in \citetalias{Kelleher:2024oip}. We find that the tension between visible baryonic masses and the dynamical mass of galaxy clusters is significantly relaxed by applying the IGIMF theory, which provides a significantly more realistic quantification of stellar populations than an invariant canonical IMF.

     We also clarify the potential tension between the baryonic masses of high-luminosity elliptical galaxies predicted by the IGIMF and previous studies in which baryonic masses derived from a canonical IMF were found to agree with MONDian dynamical or lensing masses. We suggest that this tension may arise from several factors: (1) the mass models themselves may be in conflict with recent observations of stellar populations in elliptical galaxies; (2) precise IGIMF models based on the intrinsic stellar population properties of each galaxy have not been constructed; and (3) the spatial distribution of stellar remnants may extend beyond the effective radius typically adopted as the observational aperture. A thorough assessment of this tension would require detailed modeling of both the stellar populations and the dynamics of individual galaxies, which is beyond the scope of the present work.

     It is noteworthy that the IGIMF is an independent theory describing the mass distribution of stars in galaxies, with no direct connection to MOND. Therefore, the significant alleviation of the missing mass problem under the IGIMF framework is not due to any parameter tuning or adjustments made to fit MOND to the data. This enhances the robustness and credibility of the presented results.

     We acknowledge that the ICL component adopted here is primarily qualitative. As the identification, distribution, and formation mechanisms of the ICL require further investigation, more precise models are needed to improve estimates of the ICL mass fraction.

     Additionally, our analysis is restricted to nearby galaxy clusters with redshifts below 0.1. For higher-redshift clusters, adjustments to the IGIMF $M/L$ ratios are necessary to account for their younger stellar populations. In particular, the Bullet Cluster (at $z \approx 0.3$), whose mass is constrained by lensing observations, represents an interesting target for further investigation and will be studied in future work.

     Finally, to further test this approach, we recommend selecting several galaxy clusters in hydrostatic equilibrium, spanning different redshift values, and first analyzing the BCG masses using velocity dispersion and lensing methods. These can then be compared with the stellar masses derived from the IGIMF theory. The purpose is to test the IGIMF $M/L$ ratios. If the IGIMF $M/L$ ratios are confirmed, high-precision measurements of the galaxy population, the ICL and ICM should follow, and the three independent mass analyses (velocity dispersion, lensing, X-ray gas) should be compared to ensure that any remaining missing masses are accounted for. And a final comparison can be made with the total baryonic mass predicted by the IGIMF on the galaxy cluster scale.

\section*{acknowledgements}
    The authors thank J. Pflamm-Altenburg, I. Thies, and I. Banik for valuable discussions.   
    This publication makes use of data products from the Two Micron All Sky Survey, which is a joint project of the University of Massachusetts and the Infrared Processing and Analysis Center/California Institute of Technology, funded by the National Aeronautics and Space Administration and the National Science Foundation.
    The authors would like to acknowledge the following sources of funding: D. Zhang gratefully acknowledges financial support from the China Scholarship Council (CSC). A. H. Zonoozi thanks the Alexander von Humboldt Foundation (AvH) for support. P. Kroupa acknowledges support through grant 26-21774S from the Czech Grant Agency and also through the DAAD Eastern-European Exchange Scheme between Bonn and Prague. We thank the anonymous referee for the constructive comments and helpful discussion.



\bibliography{ref.bib} 

\begin{appendix}

\section{The M/L ratios and color relations in different bands}

     \begin{figure}[H]
     \centering
     \includegraphics[width=0.45\textwidth]{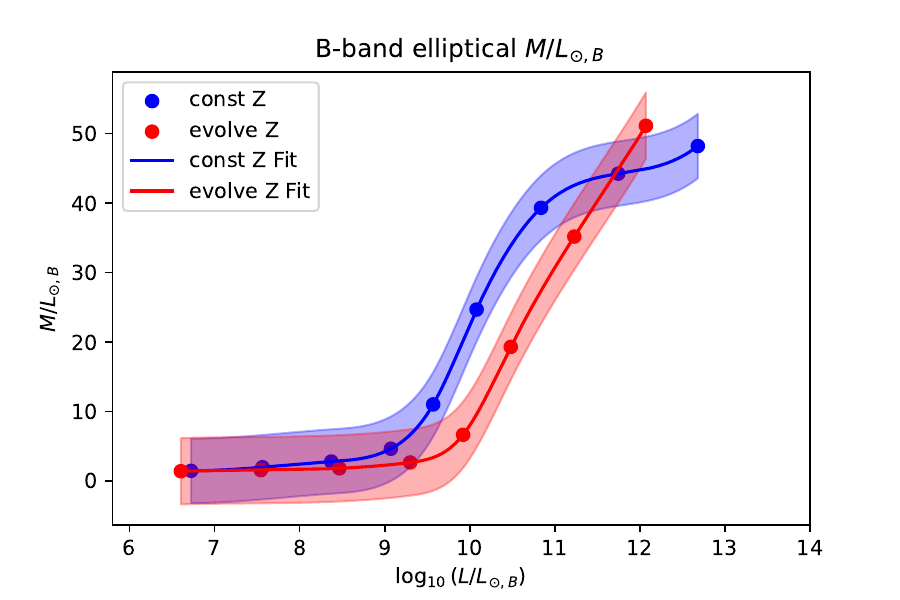}
     \includegraphics[width=0.45\textwidth]{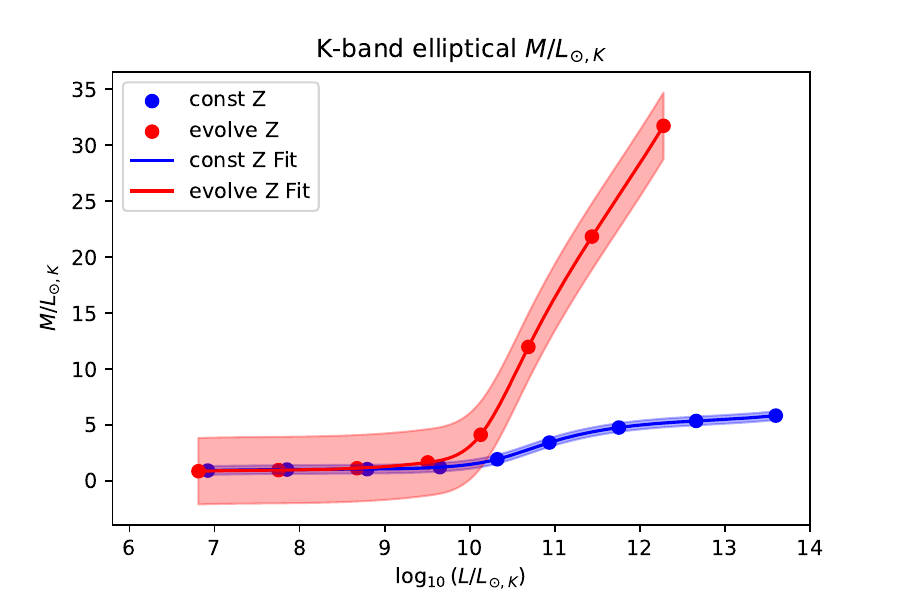}
     \caption{$M/L$ ratios of elliptical galaxies under the IGIMF framework, comparing the constant-metallicity (const Z) and evolving-metallicity (evolve Z) cases. 
     Top panel: B-band $M/L$ ratios. Blue dots indicate the values for the constant-metallicity case; the blue solid line shows the cubic spline interpolation, and the blue shaded region marks the interpolation uncertainty range (error bounds from \citet{hall1976optimal}). Red dots, line, and shaded region give the corresponding results for the evolving-metallicity case. Bottom panel: Same as the top panel, but for the K-band.}
     \label{BK_evo}
     \end{figure}

     \begin{figure}[H]
     \centering
     \includegraphics[width=0.45\textwidth]{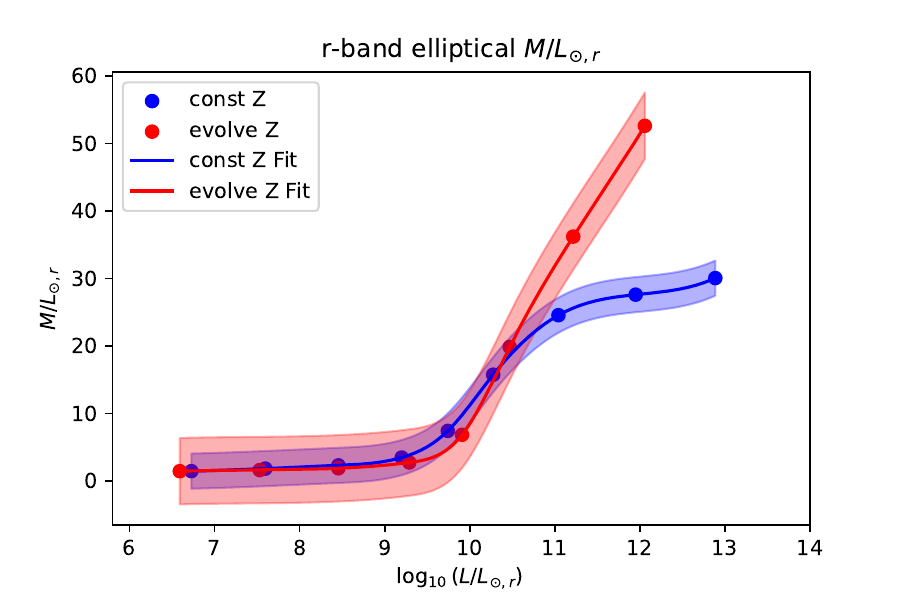}
     \includegraphics[width=0.45\textwidth]{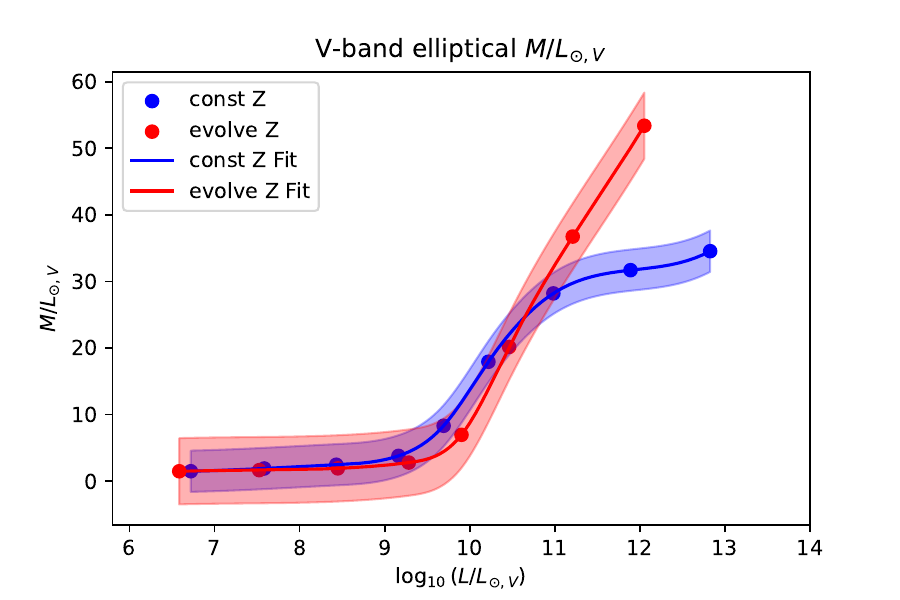}
     \caption{As described in Fig.~\ref{BK_evo}, but in the r- (top panel) and V- (bottom panel) bands.}
     \label{Vr_evo}
     \end{figure}

     \begin{figure}[H]
     \centering
     \includegraphics[width=0.45\textwidth]{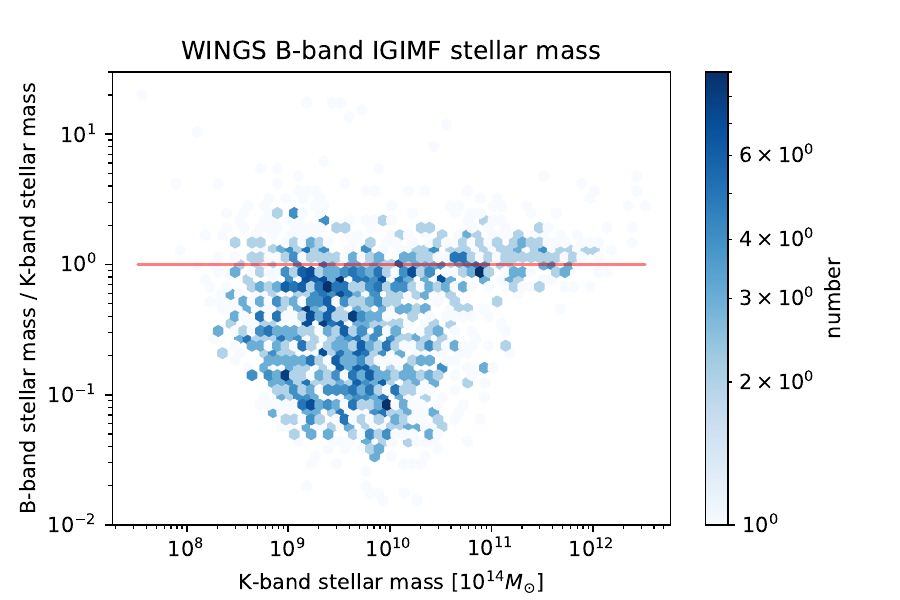}
     \includegraphics[width=0.45\textwidth]{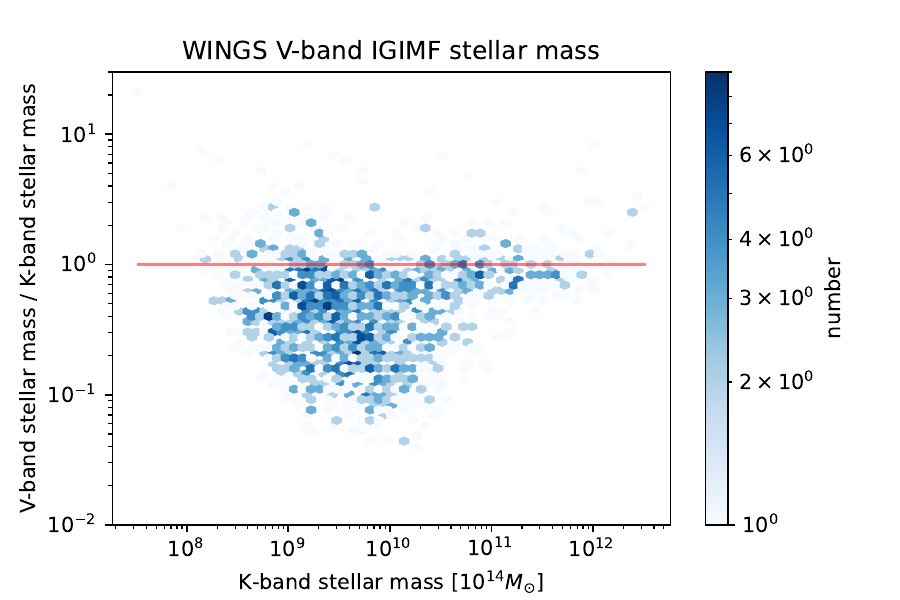}
     \caption{Comparison of IGIMF-based normalized stellar masses of a subset of WINGS member galaxies derived from the B-, V-, and K-bands. Top panel: B-band mass versus K-band mass, with the B-band masses shown as a number density map; all stellar masses are normalized to the K-band mass of each galaxy. Bottom panel: V-band mass versus K-band mass; the rest are the same as in the top panel. The red solid line in both panels indicates the normalized K-band mass of each galaxy. Data are from \cite{fasano2012morphology,d2014surface}.}
     \label{BVK_IGIMF}
     \end{figure}

     \begin{figure}[H]
     \centering
     \includegraphics[width=0.45\textwidth]{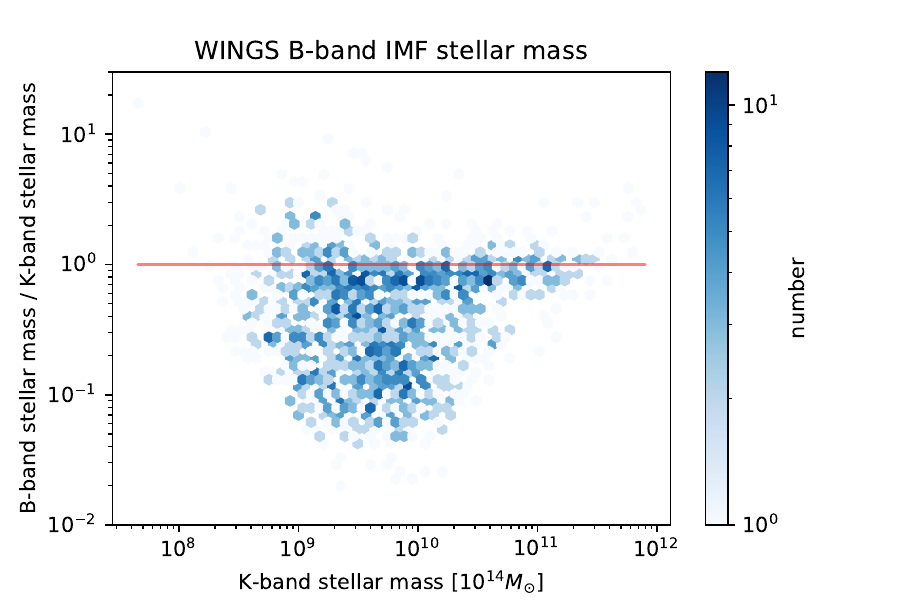}
     \includegraphics[width=0.45\textwidth]{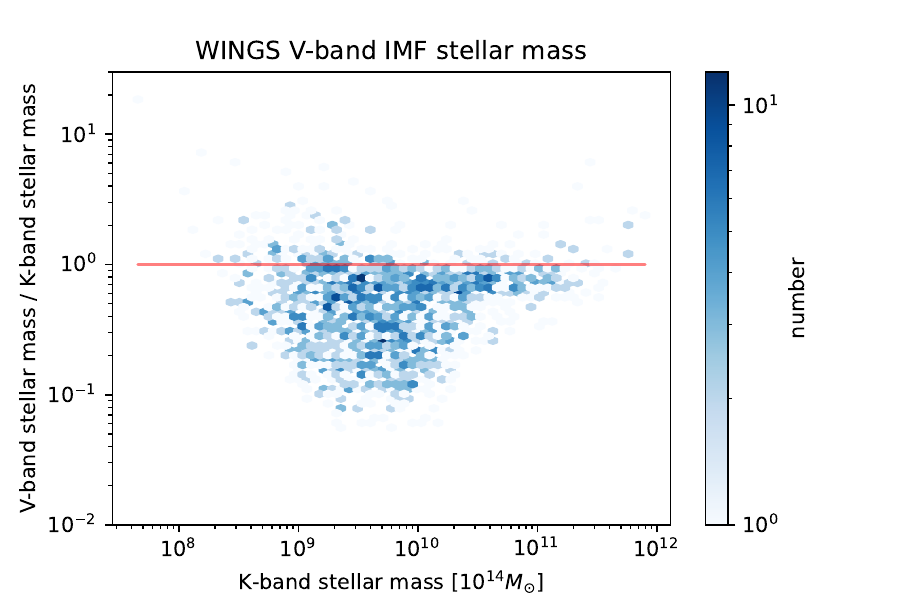}
     \caption{As described in Fig.~\ref{BVK_IGIMF}, but in the canonical IMF case.}
     \label{BVK_IMF}
     \end{figure}

     \begin{figure}[H]
     \centering
     \includegraphics[width=0.45\textwidth]{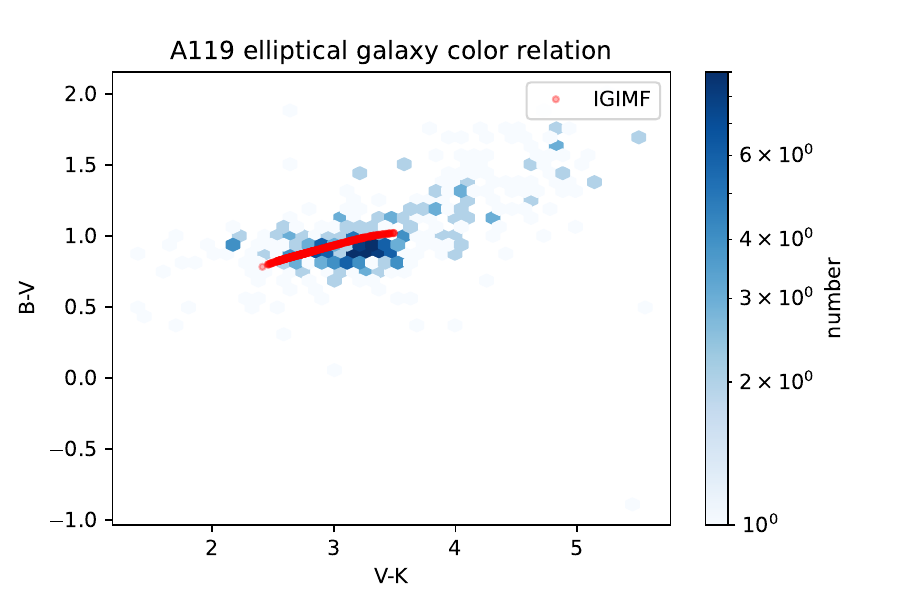}
     \includegraphics[width=0.45\textwidth]{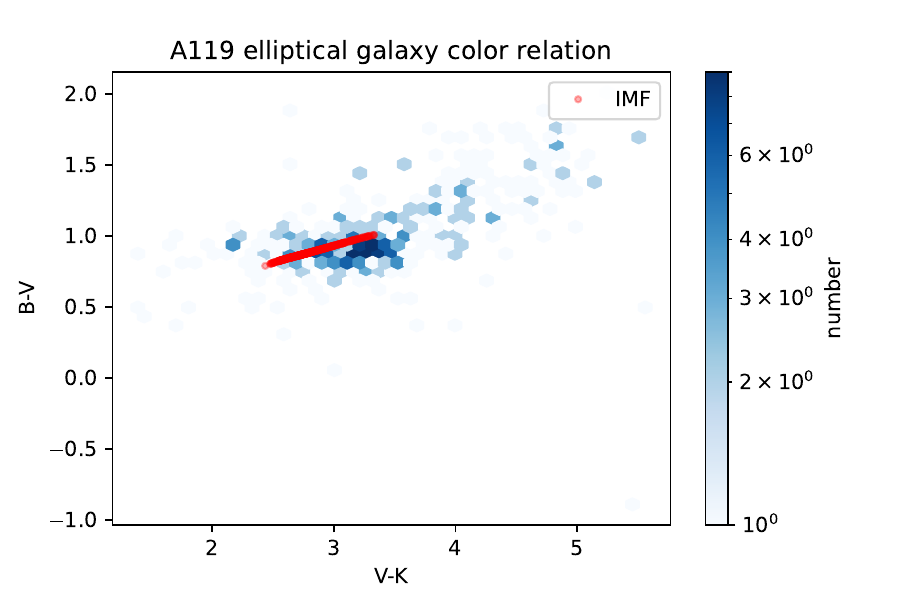}
     \caption{Color--color relations of a subset of elliptical galaxies in the Abell 119 cluster. 
     Top panel: Predicted colors under the IGIMF framework (red dots) compared with the observed relations from the WINGS data (shown as a density distribution). Bottom panel: Same as the top panel, but for the canonical IMF. The observed apparent magnitudes in the B-, V-, and K-bands are taken from \citet{d2014surface}.}
     \label{color_compare}
     \end{figure}

     \begin{figure}[H]
     \centering
     \includegraphics[width=0.45\textwidth]{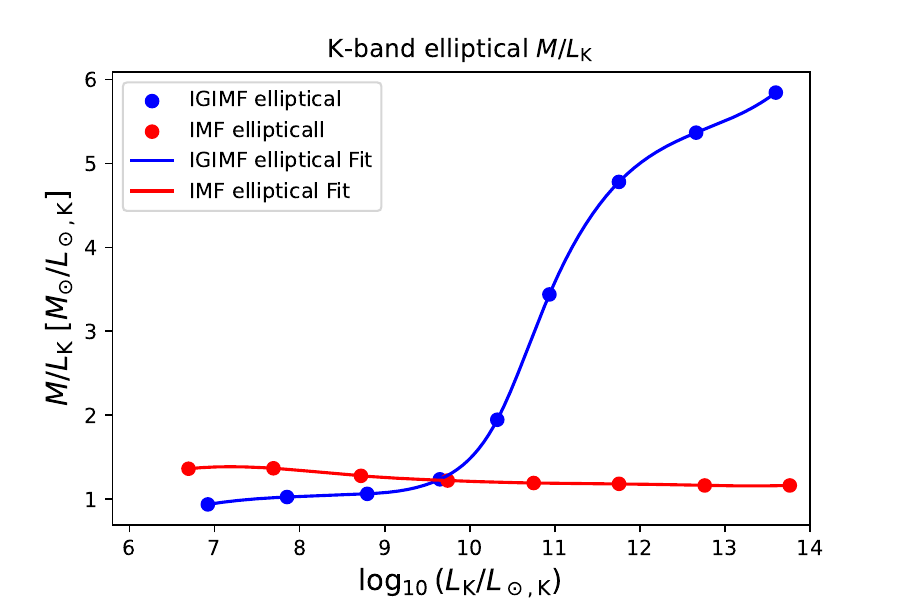}
     \includegraphics[width=0.45\textwidth]{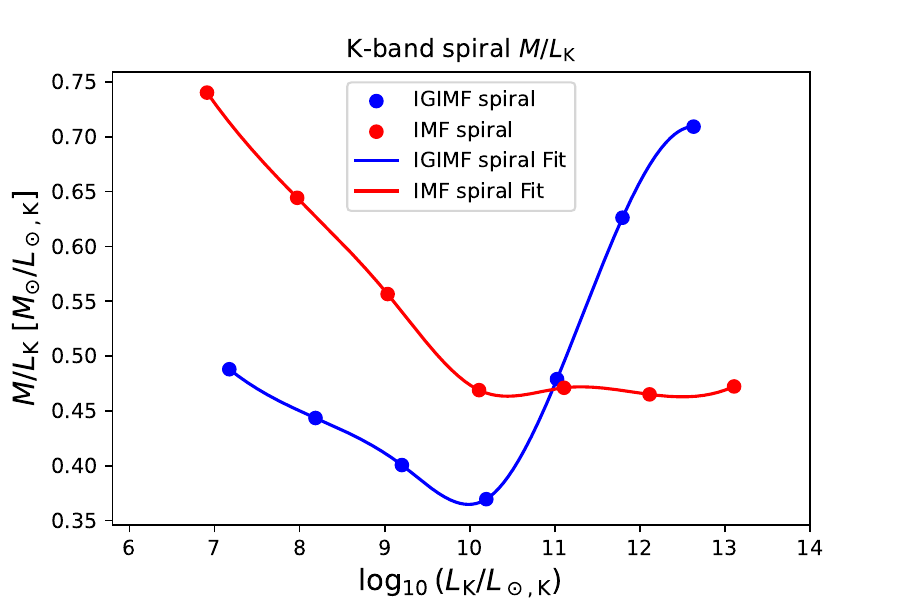}
     \caption{As described in Fig.~\ref{Bband}, but in the K-band.}
     \label{Kband}
     \end{figure}
     \begin{figure}[H]
     \centering
     \includegraphics[width=0.45\textwidth]{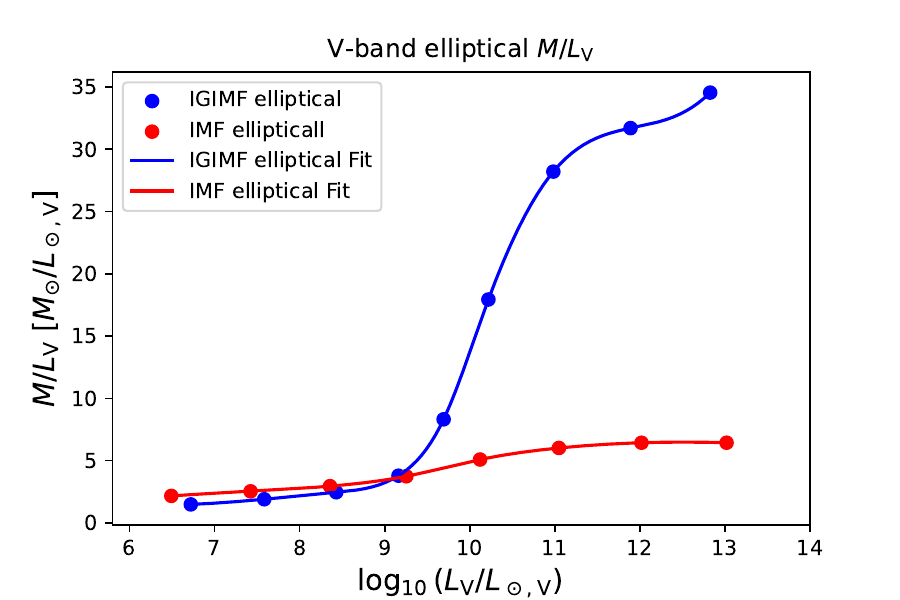}
     \includegraphics[width=0.45\textwidth]{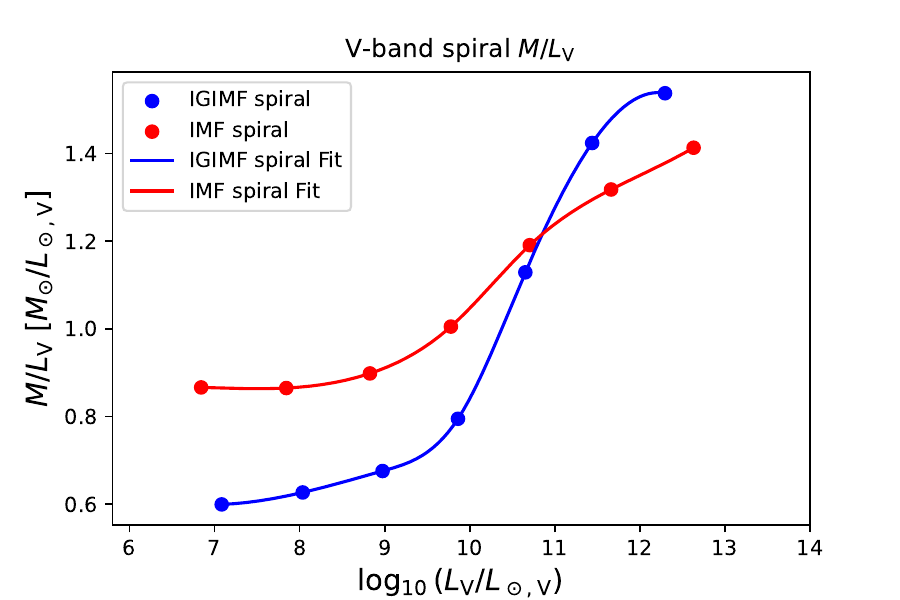}
     \caption{As described in Fig.~\ref{Bband}, but in the V-band.}
     \label{Vband}
     \end{figure}
     \begin{figure}[H]
     \centering
     \includegraphics[width=0.45\textwidth]{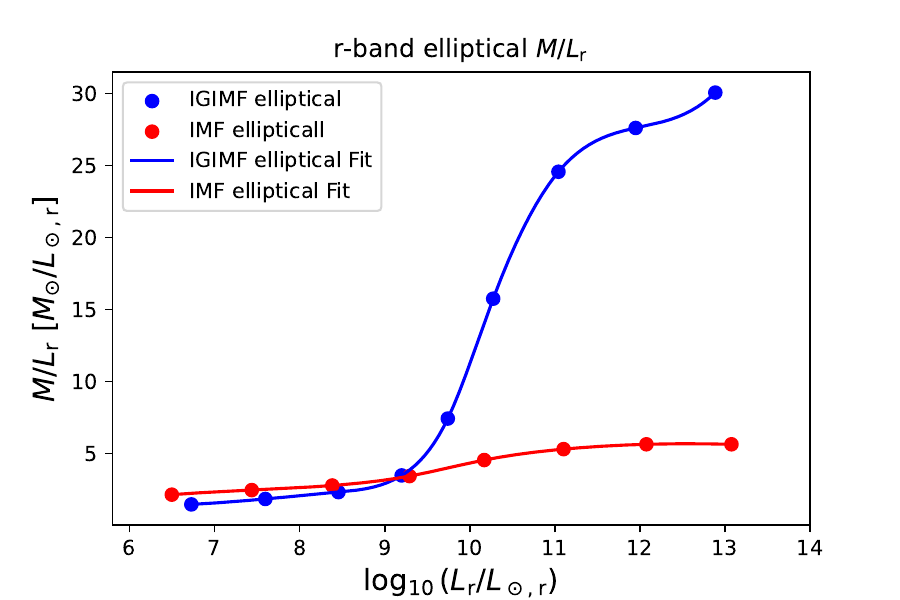}
     \includegraphics[width=0.45\textwidth]{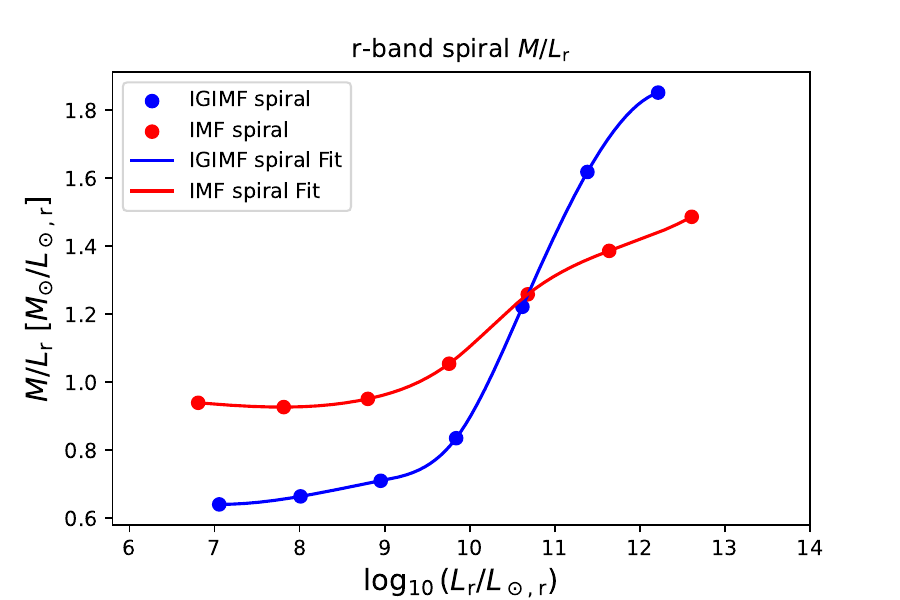}
     \caption{As described in Fig.~\ref{Bband}, but in the r-band.}
     \label{Rband}
     \end{figure}
     \begin{figure}[H]
     \centering
     \includegraphics[width=0.45\textwidth]{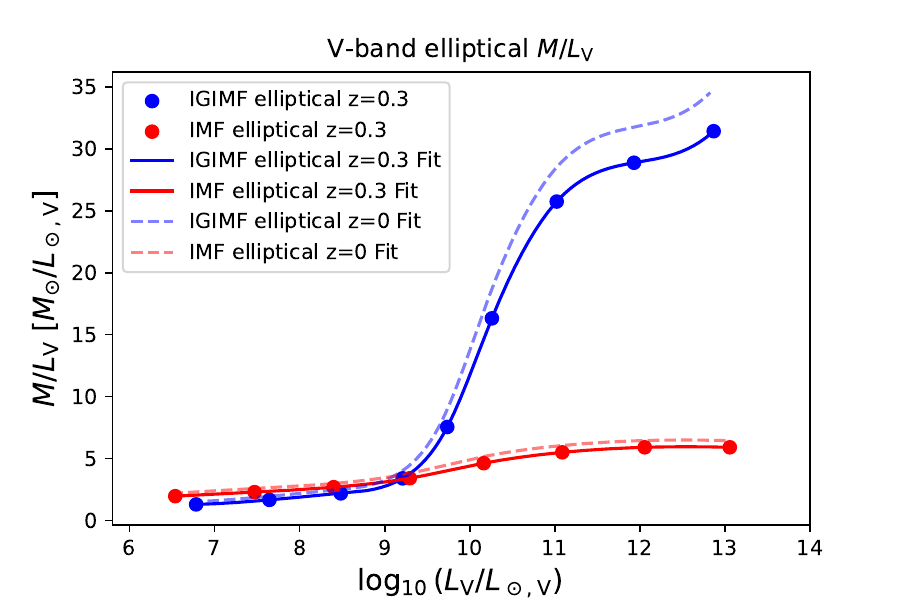}
     \caption{As described in Fig.~\ref{Bband}, but in the V-band at redshift $z = 0.3$. The dashed lines are the results from upper panel of Fig.~\ref{Vband}.}
     \label{Vband_0.3}
     \end{figure}
     \begin{figure}[H]
     \centering
     \includegraphics[width=0.45\textwidth]{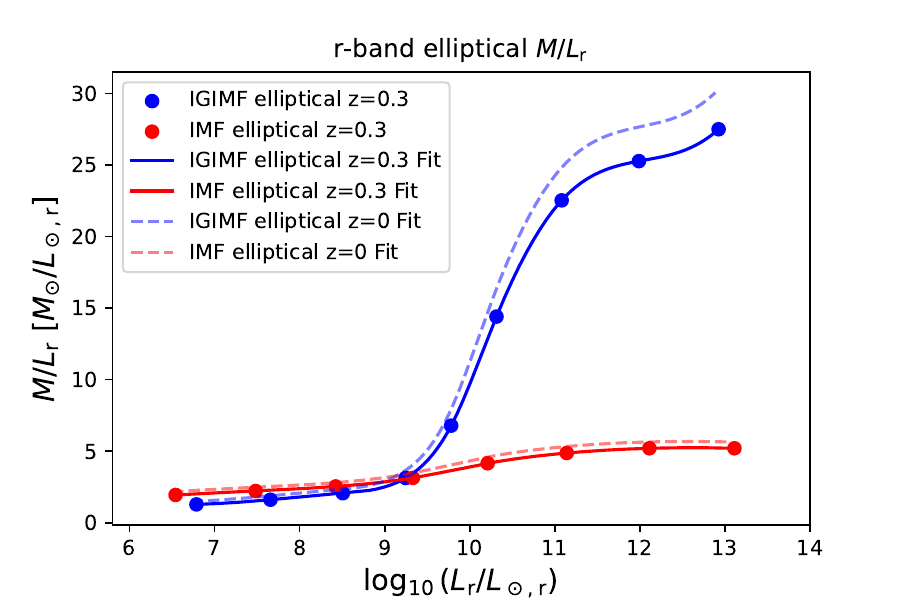}
     \caption{As described in Fig.~\ref{Bband}, but in the r-band at redshift $z = 0.3$. The dashed lines are the results from upper panel of Fig.~\ref{Rband}.}
     \label{rband_0.3}
     \end{figure}

\section{Result tables for all galaxy clusters}

     \begin{table*}
     \caption{Mass components of NGC 5044 and 12 WINGS galaxy clusters. All masses have units of $10^{14}M_{\odot}$.}            
     \label{WINGS_table}      
     \centering                          
     \begin{ruledtabular}
     \begin{tabular}{c c c c c c c c c c c c}        
     Name &  $z$ &   $M_{\rm gas}$  &  $M_{\rm g,IMF}$  &  $M_{\rm g,IG}$   & $M_{\rm I,IMF}$  & $M_{\rm I,IG}$ & $M_{\rm tot,IMF}$ & $M_{\rm tot,IG}$ & $M_{\rm M,dyn}$  & $M_{\rm N,dyn}$ &  $f_{\rm ICL}$\\    
     \hline                        
     A85 & 0.0521 & $1.48^{0.14}_{0.16}$ & $0.08^{0.00}_{0.00}$ & $0.28^{0.01}_{0.01}$ & $0.01^{0.00}_{0.00}$ & $0.03^{0.00}_{0.00}$ & $1.57^{0.14}_{0.16}$ & $1.79^{0.14}_{0.16}$ & $1.83^{0.21}_{0.21}$ & $9.02^{0.53}_{0.53}$ & $0.11^{0.01}_{0.01}$ \\
     A119 & 0.0444 & $0.73^{0.10}_{0.10}$ & $0.10^{0.00}_{0.00}$ & $0.35^{0.01}_{0.01}$ & $0.04^{0.04}_{0.02}$ & $0.15^{0.15}_{0.08}$ & $0.87^{0.11}_{0.10}$ & $1.23^{0.18}_{0.13}$ & $1.76^{0.26}_{0.25}$ & $6.88^{0.53}_{0.50}$ & $0.30^{0.20}_{0.20}$ \\
     A133 & 0.0603 & $0.37^{0.04}_{0.04}$ & $0.08^{0.00}_{0.00}$ & $0.28^{0.01}_{0.01}$ & $0.03^{0.03}_{0.02}$ & $0.12^{0.12}_{0.06}$ & $0.48^{0.05}_{0.04}$ & $0.76^{0.12}_{0.07}$ & $0.55^{0.57}_{0.26}$ & $3.13^{1.65}_{0.74}$ & $0.30^{0.20}_{0.20}$ \\
     A1795 & 0.0622 & $0.83^{0.07}_{0.09}$ & $0.14^{0.00}_{0.00}$ & $0.54^{0.03}_{0.03}$ & $0.06^{0.07}_{0.03}$ & $0.23^{0.29}_{0.12}$ & $1.03^{0.10}_{0.10}$ & $1.60^{0.30}_{0.16}$ & $2.84^{0.69}_{0.69}$ & $9.03^{1.16}_{1.16}$ & $0.30^{0.20}_{0.20}$ \\
     A2589 & 0.0416 & $0.25^{0.04}_{0.04}$ & $0.05^{0.00}_{0.00}$ & $0.16^{0.01}_{0.01}$ & $0.02^{0.02}_{0.01}$ & $0.07^{0.07}_{0.04}$ & $0.32^{0.04}_{0.04}$ & $0.48^{0.08}_{0.05}$ & $0.65^{0.76}_{0.38}$ & $2.90^{1.73}_{0.87}$ & $0.30^{0.20}_{0.20}$ \\
     A2657 & 0.04 & $0.30^{0.03}_{0.04}$ & $0.05^{0.00}_{0.00}$ & $0.18^{0.01}_{0.01}$ & $0.02^{0.02}_{0.01}$ & $0.08^{0.09}_{0.04}$ & $0.37^{0.04}_{0.04}$ & $0.56^{0.10}_{0.06}$ & $0.57^{0.09}_{0.09}$ & $2.94^{0.25}_{0.24}$ & $0.30^{0.20}_{0.20}$ \\
     A2734 & 0.0624 & $0.36^{0.09}_{0.09}$ & $0.06^{0.00}_{0.00}$ & $0.22^{0.01}_{0.01}$ & $0.03^{0.03}_{0.01}$ & $0.10^{0.09}_{0.05}$ & $0.45^{0.09}_{0.09}$ & $0.68^{0.13}_{0.10}$ & $0.76^{0.27}_{0.23}$ & $3.53^{0.63}_{0.55}$ & $0.30^{0.20}_{0.20}$ \\
     A3158 & 0.059 & $0.73^{0.13}_{0.14}$ & $0.13^{0.00}_{0.00}$ & $0.47^{0.01}_{0.01}$ & $0.06^{0.06}_{0.03}$ & $0.20^{0.20}_{0.11}$ & $0.92^{0.14}_{0.14}$ & $1.41^{0.24}_{0.18}$ & $1.89^{0.20}_{0.17}$ & $6.91^{0.39}_{0.33}$ & $0.30^{0.20}_{0.20}$ \\
     A3530 & 0.0544 & $0.23^{0.08}_{0.07}$ & $0.05^{0.00}_{0.00}$ & $0.18^{0.01}_{0.01}$ & $0.02^{0.02}_{0.01}$ & $0.08^{0.08}_{0.04}$ & $0.30^{0.08}_{0.07}$ & $0.49^{0.11}_{0.08}$ & $1.05^{0.44}_{0.34}$ & $3.56^{0.79}_{0.60}$ & $0.30^{0.20}_{0.20}$ \\
     A3532 & 0.0555 & $0.38^{0.08}_{0.08}$ & $0.09^{0.00}_{0.00}$ & $0.34^{0.01}_{0.01}$ & $0.04^{0.04}_{0.02}$ & $0.14^{0.14}_{0.08}$ & $0.51^{0.09}_{0.08}$ & $0.86^{0.16}_{0.11}$ & $1.15^{0.19}_{0.16}$ & $4.41^{0.37}_{0.32}$ & $0.30^{0.20}_{0.20}$ \\
     A4059 & 0.048 & $0.33^{0.05}_{0.05}$ & $0.07^{0.00}_{0.00}$ & $0.28^{0.01}_{0.01}$ & $0.03^{0.03}_{0.02}$ & $0.12^{0.12}_{0.06}$ & $0.43^{0.06}_{0.05}$ & $0.73^{0.13}_{0.08}$ & $0.88^{0.12}_{0.12}$ & $3.71^{0.27}_{0.27}$ & $0.30^{0.20}_{0.20}$ \\
     MKW3s & 0.0453 & $0.29^{0.03}_{0.04}$ & $0.05^{0.00}_{0.00}$ & $0.17^{0.01}_{0.01}$ & $0.02^{0.02}_{0.01}$ & $0.07^{0.07}_{0.04}$ & $0.36^{0.04}_{0.04}$ & $0.53^{0.08}_{0.06}$ & $0.62^{0.07}_{0.07}$ & $2.96^{0.17}_{0.17}$ & $0.30^{0.20}_{0.20}$ \\
     
     NGC 5044 & 0.009 & $0.015^{0.002}_{0.003}$ & $0.01^{0.00}_{0.00}$ & $0.03^{0.00}_{0.00}$ & $0.00^{0.00}_{0.00}$ & $0.01^{0.00}_{0.00}$ & $0.03^{0.00}_{0.00}$ & $0.05^{0.00}_{0.00}$ & $0.04^{0.00}_{0.00}$ & $0.30^{0.00}_{0.00}$ & $0.22^{0.03}_{0.03}$ \\    
     \end{tabular}
     \end{ruledtabular}
     \footnotesize{From left to right: Name of cluster; redshift $z$, NGC 5044's values from \citet{zhang2011hiflugcs}, rest from \citet{fasano2006wings}; gas mass $M_{\rm gas}$, within virial radius, all from \citet{brownstein2006galaxy}; galaxy mass $M_{\rm g,IMF}$, assuming the canonical IMF; galaxy mass $M_{\rm g,IG}$, assuming the IGIMF; ICL mass $M_{\rm I,IMF}$, assuming the canonical IMF; ICL mass $M_{\rm I,IG}$, assuming the IGIMF; total baryonic mass $M_{\rm tot,IMF}$, assuming the canonical IMF; total baryonic mass $M_{\rm tot,IG}$, assuming the IGIMF; the MOND dynamical masses $M_{\rm M,dyn}$, from \citet{brownstein2006galaxy}; the Newtonian dynamical masses $M_{\rm N,dyn}$, from \citet{brownstein2006galaxy}; the ICL luminosity fraction is $f_{\rm ICL}$.     
     The ICL fraction of A85 and NGC 5044 from \citet{montes2021buildup}, and \citet{ragusa2023does} respectively. The error bars of $M_{\rm I,IG}$, $M_{\rm I,IMF}$, $M_{\rm tot,IG}$, and $M_{\rm tot,IMF}$ all include the systematic uncertainty of the ICL mass. }
     \end{table*}

     \begin{table*}
     \caption{Mass components of 33 2MASS galaxy clusters. All masses have units of $10^{14}M_{\odot}$.}             
     \label{2MASS_table}      
     \centering                          
     \begin{ruledtabular}
     \begin{tabular}{c c c c c c c c c c c c}        
     Name &  $z$ &  $M_{\rm gas}$  &  $M_{\rm g,IMF}$  &  $M_{\rm g,IG}$   & $M_{\rm I,IMF}$  & $M_{\rm I,IG}$ & $M_{\rm tot,IMF}$  & $M_{\rm tot,IG}$   & $M_{\rm M,dyn}$  & $M_{\rm N,dyn}$ &  $f_{\rm ICL}$\\    
     \hline                        
     A2319 & 0.0557 & $2.66^{0.35}_{0.39}$ & $0.17^{0.01}_{0.01}$ & $0.42^{0.02}_{0.02}$ & $0.07^{0.10}_{0.05}$ & $0.18^{0.24}_{0.13}$ & $2.91^{0.36}_{0.39}$ & $3.26^{0.43}_{0.41}$ & $3.60^{0.45}_{0.45}$ & $15.07^{0.98}_{0.96}$ & $0.30^{0.20}_{0.20}$ \\
     A2029 & 0.0773 & $1.55^{0.14}_{0.17}$ & $0.11^{0.01}_{0.01}$ & $0.28^{0.03}_{0.03}$ & $0.06^{0.00}_{0.00}$ & $0.14^{0.00}_{0.00}$ & $1.71^{0.14}_{0.17}$ & $1.97^{0.14}_{0.17}$ & $3.71^{0.78}_{0.78}$ & $12.77^{1.41}_{1.41}$ & $0.34^{0.00}_{0.00}$ \\
     A2142 & 0.0899 & $2.39^{0.26}_{0.30}$ & $0.11^{0.01}_{0.01}$ & $0.27^{0.03}_{0.03}$ & $0.05^{0.06}_{0.04}$ & $0.11^{0.15}_{0.09}$ & $2.55^{0.27}_{0.30}$ & $2.77^{0.30}_{0.31}$ & $4.36^{1.30}_{0.96}$ & $15.92^{2.47}_{1.82}$ & $0.30^{0.20}_{0.20}$ \\
     A0478 & 0.0881 & $1.30^{0.12}_{0.15}$ & $0.13^{0.02}_{0.02}$ & $0.33^{0.05}_{0.05}$ & $0.06^{0.08}_{0.04}$ & $0.14^{0.19}_{0.11}$ & $1.49^{0.14}_{0.16}$ & $1.77^{0.23}_{0.19}$ & $3.49^{0.64}_{1.11}$ & $11.45^{1.10}_{1.91}$ & $0.30^{0.20}_{0.20}$ \\
     A1651 & 0.086 & $0.81^{0.12}_{0.13}$ & $0.11^{0.02}_{0.02}$ & $0.26^{0.05}_{0.05}$ & $0.02^{0.00}_{0.00}$ & $0.04^{0.00}_{0.00}$ & $0.94^{0.12}_{0.13}$ & $1.10^{0.13}_{0.14}$ & $2.03^{0.28}_{0.28}$ & $7.38^{0.53}_{0.53}$ & $0.13^{0.00}_{0.00}$ \\
     A3391 & 0.0531 & $0.51^{0.11}_{0.11}$ & $0.08^{0.00}_{0.00}$ & $0.22^{0.02}_{0.02}$ & $0.03^{0.04}_{0.02}$ & $0.09^{0.12}_{0.07}$ & $0.62^{0.12}_{0.11}$ & $0.82^{0.17}_{0.13}$ & $1.29^{0.32}_{0.31}$ & $5.28^{0.68}_{0.66}$ & $0.30^{0.20}_{0.20}$ \\
     A3822 & 0.076 & $0.67^{0.57}_{0.52}$ & $0.08^{0.01}_{0.01}$ & $0.22^{0.02}_{0.02}$ & $0.04^{0.05}_{0.03}$ & $0.10^{0.13}_{0.07}$ & $0.79^{0.57}_{0.52}$ & $0.99^{0.58}_{0.53}$ & $1.27^{0.91}_{0.62}$ & $5.63^{2.07}_{1.41}$ & $0.30^{0.20}_{0.20}$ \\
     A3112 & 0.075 & $0.64^{0.08}_{0.09}$ & $0.06^{0.01}_{0.01}$ & $0.18^{0.02}_{0.02}$ & $0.03^{0.04}_{0.02}$ & $0.08^{0.10}_{0.06}$ & $0.73^{0.09}_{0.09}$ & $0.89^{0.13}_{0.11}$ & $1.26^{0.33}_{0.46}$ & $5.50^{0.73}_{1.04}$ & $0.30^{0.20}_{0.20}$ \\
     A2199 & 0.0303 & $0.36^{0.07}_{0.07}$ & $0.05^{0.00}_{0.00}$ & $0.13^{0.01}_{0.01}$ & $0.02^{0.03}_{0.02}$ & $0.05^{0.07}_{0.04}$ & $0.43^{0.08}_{0.07}$ & $0.54^{0.10}_{0.08}$ & $0.96^{0.08}_{0.09}$ & $3.81^{0.17}_{0.19}$ & $0.30^{0.20}_{0.20}$ \\
     A2063 & 0.0355 & $0.33^{0.05}_{0.05}$ & $0.04^{0.00}_{0.00}$ & $0.08^{0.01}_{0.01}$ & $0.02^{0.02}_{0.01}$ & $0.03^{0.05}_{0.03}$ & $0.38^{0.05}_{0.05}$ & $0.45^{0.07}_{0.06}$ & $0.58^{0.05}_{0.05}$ & $3.03^{0.12}_{0.12}$ & $0.30^{0.20}_{0.20}$ \\
     A0496 & 0.0328 & $0.74^{0.06}_{0.07}$ & $0.05^{0.00}_{0.00}$ & $0.12^{0.01}_{0.01}$ & $0.02^{0.03}_{0.01}$ & $0.05^{0.07}_{0.04}$ & $0.80^{0.07}_{0.07}$ & $0.91^{0.09}_{0.08}$ & $0.55^{0.02}_{0.02}$ & $4.01^{0.09}_{0.09}$ & $0.30^{0.20}_{0.20}$ \\
     AWM7 & 0.0172 & $0.22^{0.05}_{0.05}$ & $0.03^{0.00}_{0.00}$ & $0.08^{0.01}_{0.01}$ & $0.01^{0.02}_{0.01}$ & $0.03^{0.04}_{0.02}$ & $0.26^{0.05}_{0.05}$ & $0.33^{0.07}_{0.06}$ & $0.83^{0.10}_{0.09}$ & $3.05^{0.19}_{0.18}$ & $0.30^{0.20}_{0.20}$ \\
     A2634 & 0.0314 & $0.25^{0.06}_{0.06}$ & $0.05^{0.00}_{0.00}$ & $0.14^{0.01}_{0.01}$ & $0.02^{0.03}_{0.02}$ & $0.06^{0.08}_{0.05}$ & $0.32^{0.07}_{0.06}$ & $0.46^{0.10}_{0.08}$ & $0.69^{0.18}_{0.16}$ & $3.05^{0.41}_{0.37}$ & $0.30^{0.20}_{0.20}$ \\
     2A0335 & 0.0349 & $0.33^{0.03}_{0.04}$ & $0.04^{0.00}_{0.00}$ & $0.10^{0.01}_{0.01}$ & $0.02^{0.02}_{0.01}$ & $0.04^{0.06}_{0.03}$ & $0.39^{0.04}_{0.04}$ & $0.47^{0.06}_{0.05}$ & $0.41^{0.02}_{0.02}$ & $2.51^{0.06}_{0.06}$ & $0.30^{0.20}_{0.20}$ \\
     A3526 & 0.0114 & $0.20^{0.05}_{0.04}$ & $0.04^{0.00}_{0.00}$ & $0.11^{0.01}_{0.01}$ & $0.02^{0.02}_{0.01}$ & $0.05^{0.06}_{0.03}$ & $0.25^{0.05}_{0.04}$ & $0.36^{0.08}_{0.05}$ & $0.45^{0.03}_{0.03}$ & $2.35^{0.08}_{0.08}$ & $0.30^{0.20}_{0.20}$ \\
     A1367 & 0.0216 & $0.27^{0.04}_{0.04}$ & $0.04^{0.00}_{0.00}$ & $0.09^{0.01}_{0.01}$ & $0.02^{0.02}_{0.01}$ & $0.04^{0.05}_{0.03}$ & $0.32^{0.05}_{0.04}$ & $0.40^{0.07}_{0.05}$ & $0.75^{0.11}_{0.10}$ & $3.19^{0.24}_{0.22}$ & $0.30^{0.20}_{0.20}$ \\
     HCG094 & 0.0417 & $0.24^{0.02}_{0.03}$ & $0.03^{0.00}_{0.00}$ & $0.08^{0.01}_{0.01}$ & $0.02^{0.00}_{0.00}$ & $0.06^{0.00}_{0.00}$ & $0.29^{0.02}_{0.03}$ & $0.38^{0.02}_{0.03}$ & $0.43^{0.07}_{0.07}$ & $2.40^{0.21}_{0.21}$ & $0.42^{0.00}_{0.00}$ \\
     MKW08 & 0.027 & $0.11^{0.11}_{0.09}$ & $0.02^{0.00}_{0.00}$ & $0.05^{0.01}_{0.01}$ & $0.01^{0.01}_{0.01}$ & $0.02^{0.03}_{0.01}$ & $0.14^{0.11}_{0.09}$ & $0.18^{0.11}_{0.09}$ & $0.38^{0.21}_{0.13}$ & $1.79^{0.50}_{0.32}$ & $0.30^{0.20}_{0.20}$ \\
     A4038 & 0.0283 & $0.28^{0.04}_{0.04}$ & $0.03^{0.00}_{0.00}$ & $0.07^{0.01}_{0.01}$ & $0.01^{0.02}_{0.01}$ & $0.03^{0.04}_{0.02}$ & $0.32^{0.04}_{0.04}$ & $0.37^{0.06}_{0.05}$ & $0.40^{0.02}_{0.02}$ & $2.38^{0.06}_{0.05}$ & $0.30^{0.20}_{0.20}$ \\
     A1060 & 0.0114 & $0.07^{0.02}_{0.02}$ & $0.02^{0.00}_{0.00}$ & $0.06^{0.01}_{0.01}$ & $0.01^{0.01}_{0.01}$ & $0.02^{0.03}_{0.02}$ & $0.10^{0.02}_{0.02}$ & $0.15^{0.04}_{0.03}$ & $0.50^{0.09}_{0.08}$ & $1.69^{0.16}_{0.14}$ & $0.30^{0.20}_{0.20}$ \\
     A2052 & 0.0348 & $0.34^{0.04}_{0.04}$ & $0.04^{0.00}_{0.00}$ & $0.10^{0.01}_{0.01}$ & $0.02^{0.02}_{0.01}$ & $0.04^{0.06}_{0.03}$ & $0.39^{0.05}_{0.04}$ & $0.48^{0.07}_{0.05}$ & $0.35^{0.01}_{0.01}$ & $2.40^{0.05}_{0.05}$ & $0.30^{0.20}_{0.20}$ \\
     A0539 & 0.0288 & $0.23^{0.04}_{0.04}$ & $0.04^{0.00}_{0.00}$ & $0.09^{0.01}_{0.01}$ & $0.02^{0.02}_{0.01}$ & $0.04^{0.05}_{0.03}$ & $0.29^{0.05}_{0.04}$ & $0.36^{0.07}_{0.05}$ & $0.44^{0.05}_{0.05}$ & $2.36^{0.14}_{0.13}$ & $0.30^{0.20}_{0.20}$ \\
     AS1101 & 0.058 & $0.20^{0.02}_{0.03}$ & $0.03^{0.00}_{0.00}$ & $0.08^{0.01}_{0.01}$ & $0.01^{0.02}_{0.01}$ & $0.03^{0.05}_{0.03}$ & $0.24^{0.03}_{0.03}$ & $0.32^{0.05}_{0.04}$ & $0.50^{0.39}_{0.23}$ & $2.23^{0.89}_{0.52}$ & $0.30^{0.20}_{0.20}$ \\
     AWM4 & 0.0326 & $0.06^{0.00}_{0.00}$ & $0.02^{0.00}_{0.00}$ & $0.04^{0.01}_{0.01}$ & $0.01^{0.01}_{0.00}$ & $0.02^{0.03}_{0.01}$ & $0.08^{0.01}_{0.01}$ & $0.12^{0.03}_{0.01}$ & $0.25^{0.00}_{0.00}$ & $0.74^{0.06}_{0.06}$ & $0.30^{0.20}_{0.20}$ \\
     EXO0422 & 0.039 & $0.15^{0.09}_{0.09}$ & $0.02^{0.00}_{0.00}$ & $0.04^{0.01}_{0.01}$ & $0.01^{0.01}_{0.01}$ & $0.02^{0.02}_{0.01}$ & $0.17^{0.09}_{0.09}$ & $0.21^{0.09}_{0.09}$ & $0.57^{0.41}_{0.27}$ & $2.12^{0.79}_{0.53}$ & $0.30^{0.20}_{0.20}$ \\
     A0400 & 0.024 & $0.15^{0.02}_{0.02}$ & $0.03^{0.00}_{0.00}$ & $0.07^{0.01}_{0.01}$ & $0.01^{0.02}_{0.01}$ & $0.03^{0.04}_{0.02}$ & $0.19^{0.03}_{0.02}$ & $0.25^{0.05}_{0.03}$ & $0.20^{0.03}_{0.03}$ & $1.42^{0.10}_{0.10}$ & $0.30^{0.20}_{0.20}$ \\
     A0262 & 0.0161 & $0.26^{0.11}_{0.10}$ & $0.02^{0.00}_{0.00}$ & $0.06^{0.01}_{0.01}$ & $0.01^{0.01}_{0.01}$ & $0.02^{0.03}_{0.02}$ & $0.29^{0.11}_{0.10}$ & $0.34^{0.11}_{0.10}$ & $0.13^{0.02}_{0.02}$ & $1.39^{0.09}_{0.08}$ & $0.30^{0.20}_{0.20}$ \\
     A2151 & 0.0369 & $0.12^{0.02}_{0.02}$ & $0.04^{0.00}_{0.00}$ & $0.11^{0.01}_{0.01}$ & $0.02^{0.02}_{0.01}$ & $0.05^{0.06}_{0.04}$ & $0.18^{0.03}_{0.02}$ & $0.28^{0.07}_{0.04}$ & $0.25^{0.02}_{0.02}$ & $1.42^{0.06}_{0.06}$ & $0.30^{0.20}_{0.20}$ \\
     AS0636 & 0.0116 & $0.06^{0.04}_{0.02}$ & $0.01^{0.00}_{0.00}$ & $0.02^{0.01}_{0.01}$ & $0.01^{0.01}_{0.00}$ & $0.01^{0.01}_{0.01}$ & $0.08^{0.04}_{0.02}$ & $0.09^{0.04}_{0.02}$ & $0.09^{0.08}_{0.05}$ & $0.65^{0.29}_{0.18}$ & $0.30^{0.20}_{0.20}$ \\
     A3581 & 0.023 & $0.08^{0.02}_{0.02}$ & $0.01^{0.00}_{0.00}$ & $0.02^{0.00}_{0.00}$ & $0.00^{0.00}_{0.00}$ & $0.01^{0.01}_{0.01}$ & $0.09^{0.02}_{0.02}$ & $0.10^{0.02}_{0.02}$ & $0.14^{0.02}_{0.02}$ & $0.92^{0.06}_{0.06}$ & $0.30^{0.20}_{0.20}$ \\
     MKW04 & 0.02 & $0.09^{0.01}_{0.01}$ & $0.01^{0.00}_{0.00}$ & $0.03^{0.01}_{0.01}$ & $0.00^{0.01}_{0.00}$ & $0.01^{0.02}_{0.01}$ & $0.11^{0.01}_{0.01}$ & $0.14^{0.02}_{0.02}$ & $0.08^{0.01}_{0.01}$ & $0.78^{0.04}_{0.04}$ & $0.30^{0.20}_{0.20}$ \\
     NGC2563 & 0.0163 & $0.01^{0.00}_{0.00}$ & $0.01^{0.00}_{0.00}$ & $0.01^{0.00}_{0.00}$ & $0.00^{0.00}_{0.00}$ & $0.01^{0.01}_{0.00}$ & $0.01^{0.00}_{0.00}$ & $0.03^{0.01}_{0.01}$ & $0.02^{0.00}_{0.00}$ & $0.09^{0.01}_{0.01}$ & $0.30^{0.20}_{0.20}$ \\
     Triangul & 0.051 & $1.98^{0.22}_{0.25}$ & $0.14^{0.01}_{0.01}$ & $0.38^{0.02}_{0.02}$ & $0.06^{0.08}_{0.05}$ & $0.16^{0.22}_{0.12}$ & $2.18^{0.23}_{0.25}$ & $2.52^{0.31}_{0.28}$ & $4.48^{0.57}_{0.57}$ & $15.22^{1.01}_{1.01}$ & $0.30^{0.20}_{0.20}$ \\
     \end{tabular}
     \end{ruledtabular}
     \footnotesize{From left to right: Name of cluster; redshift $z$, from \citet{lin2004k}; gas mass $M_{\rm gas}$, within virial radius, NGC 2563 and AWM4's values from \citet{angus2008x}, rest from \citet{brownstein2006galaxy}; galaxy mass $M_{\rm g,IMF}$, assuming the canonical IMF; galaxy mass $M_{\rm g,IG}$, assuming the IGIMF; the ICL mass $M_{\rm I,IMF}$, assuming the canonical IMF; the ICL mass $M_{\rm I,IG}$, assuming the IGIMF; total baryonic mass $M_{\rm tot,IMF}$, assuming the canonical IMF; total baryonic mass $M_{\rm tot,IG}$, assuming the IGIMF; the MOND dynamical masses $M_{\rm M,dyn}$, NGC 2563 and AWM4's values from \citet{angus2008x}, the rest from \citet{brownstein2006galaxy}; the Newtonian dynamical masses $M_{\rm N,dyn}$, NGC 2563 and AWM4's values from \citet{angus2008x}, the rest from \citet{brownstein2006galaxy}; the ICL luminosity fraction is $f_{\rm ICL}$.      
     The ICL fraction of A2029 and A1651 from \citet{tang2018investigation}, and HCG94 from \citet{ragusa2021vegas}. The error bars of $M_{\rm I,IG}$, $M_{\rm I,IMF}$, $M_{\rm tot,IG}$, and $M_{\rm tot,IMF}$ all include the systematic uncertainty of the ICL mass.} 
     \end{table*}

\end{appendix}

\end{document}